\documentclass[12pt]{article}

\usepackage[a4paper,margin=1in]{geometry}
\usepackage{amsmath,amssymb,amsfonts}
\usepackage{authblk}
\usepackage{setspace}
\usepackage{cite}
\usepackage{hyperref}
\usepackage{bm}
\usepackage{graphicx}
\usepackage{subcaption}
\usepackage{float}

\title{\textbf{Criticality in Neural Network Function Space through Wilsonian Fixed Points and Finite-Width Corrections}}

\author[1,2]{Eric Howard}
\author[1]{Iftekher S. Chowdhury}
\author[2]{Hardique Dasore}
\author[3]{Hom Nath Dhungana}

\affil[1]{School of Mathematical and Physical Sciences, Macquarie University, Sydney, NSW 2000, Australia}
\affil[2]{Southern Cross Institute, Sydney, NSW, Australia}
\affil[3]{Charles Sturt University, Sydney, NSW, Australia}

\date{}

\begin{document}

\maketitle

\begin{abstract}
Neural networks can be studied not only as parameterized computational models but also as probability distributions over functions. In this paper we develop a Wilsonian interpretation of criticality in the neural network-quantum field theory correspondence, treating the infinite width Gaussian-process limit as a free field fixed point and finite width corrections as perturbations that create non-Gaussian interactions. In this way, we do not see the departure from infinite width as a small approximation error but as the process by which interaction, complexity, expressivity and phase-like behaviour enter neural network function space. Width, depth, activation nonlinearity, initialization variance, and training dynamics are considered as control parameters that change the effective action of the network ensemble. The critical regime is when the higher-order connected correlation functions become non-negligible, and when the finite width operators become relevant or marginal scaling factors, and when the function distribution becomes sensitive to scale-dependent structure. There we can view overparameterization as suppressing a relationship of interacting terms and we find that the critical structure of finite neural networks is given by finite-width effects. This framework is a theoretical basis for studying trainability, generalization, and architectural universality of neural network functions through Wilsonian fixed points, perturbations, and critical surfaces.
\end{abstract}


\section{Introduction}

Neural networks are typically described as parameterized computational systems whose behaviour is determined by weights, biases, nonlinear activation functions, architectural depth and the optimization procedure when training. This is a powerful definition because it directly relates to the real world: a neural network is built from layers, each layer has trainable parameters and learning is done by modifying their parameters so as to reduce a loss function. In this sense most of the work in deep learning theory has focused on parameters and loss landscapes, the role of gradient descent, the effect of overparameterization, the stability of the optimization and the reason why large models are generalizable even when they have many more parameters than training examples. These are essential questions to be addressed, but they do not entirely cover the geometry of neural networks. A neural network can be studied from a more abstract point of view: not as a point in parameter space alone, but as an object that generates a probability distribution over functions. This function space perspective is important, as the output of a neural network is not the parameter vector itself, but the function that connects input and output. Architecture, initialization, width, depth, and training are not just a computational graph; they are an ensemble of possible functions with some statistical structure.

The function-space view is most natural when the parameters of a neural network are randomly initialized. Before training, each draw of the weights and biases creates a different function. The set of all such possible functions defines a probability distribution over the space of input-output maps. In the infinite-width limit of the neural network, it is known that many neural network architectures converge to Gaussian models. In this case, their induced distribution over functions becomes Gaussian and is fully determined by a two-point covariance kernel. This has been one of the most important results of the theory of wide neural networks as it enables the analysis of neural networks in terms of kernels, covariance functions, and Bayesian nonparametric inference. The Gaussian process limit is very simple in that all higher-order moments are associated with the two-point function by the Gaussian factorization. In this case, the neural network prior becomes analytically tractable and the complicated description of the parameter space collapses into a simpler description of the function space. But this also implies the description of the network with finite width dimension, where there exist non-Gaussian effects and higher-order correlations are no longer present.

The correspondence between neural network and quantum field theory can be a very powerful mechanism to answer this problem. In the correspondence, the output of a neural network is treated in the sense of a field defined over the input domain and the probability distribution over the network functions is described by an effective action. The infinite-width limit of Gaussian process is then similar to a free field theory since the effective action is quadratic and the theory is described by the two-point function. The covariance kernel of the Gaussian process is analogous to a propagator and higher-order correlation functions factorize in the same way as in a free quantum field theory. It provides a precise mathematical language which can be used to study the statistical structure of neural network function space with the correlation functions, effective actions, perturbative corrections and renormalization-group flow. The correspondence thus allows neural networks to be studied in terms of concepts originally developed for field theories, statistical mechanics, and critical phenomena \cite{halverson2021,grosvenor2022,erbin2021,erbin2022}.

Another line of work has connected deep learning architectures to renormalization-group coarse-graining in a holographic setting. To identify the relevant physical degrees of freedom, to induce scale-invariant representations, and to relate the learned RG-like maps to holographic entanglement and the AdS/CFT correspondence, a new information-theoretic neural network architecture has been proposed. The holographic renormalization and learned coarse-graining are discussed in this work, while we focus on the effective function-space theory of the infinite- and finite-width neural networks.

The most important aspect of this correspondence for this paper is the interpretation of finite-width effects. A neural network with finite width is not exactly Gaussian in function space. Its higher-order connected correlation functions do not always disappear and the induced probability distribution cannot always be described by a purely quadratic effective action. In field theory, this means that finite-width corrections create interactions. These interactions occur as non-Gaussian terms in the effective action and as nonzero connected higher-order correlation functions in the function space distribution. A finite-width neural network could therefore be thought of as an interacting effective field theory rather than as a very imperfect approximation to an infinite-width Gaussian process. This is a key point. If finite-width corrections are treated only as technical errors around the infinite-width limit, their significance is minimized. If they are considered as interaction terms in an effective theory, then they are a possibility for expressivity, complexity, instability, and critical behaviour \cite{halverson2021,erbin2021,grosvenor2022}.

In this paper, we propose a Wilsonian interpretation of criticality in the neural network-quantum field theory correspondence. The basic idea is that the infinite-width Gaussian-process limit is a fixed point in neural network function space and finite-width corrections can be viewed as perturbations around such a fixed point. In the Wilsonian analysis of field theory, a fixed point is not a static solution but rather a theory whose form is not changed with the scale. Perturbations around a fixed point are classified according to whether they grow, decay, or remain marginal in the flow of renormalization groups. Relevant perturbations grow and explain the behaviour of the network on a large scale, irrelevant perturbations decay and become irrelevant on a large scale, and marginal perturbations are at the boundary and can influence the critical or near-critical behaviour. By applying this logic to neural network function space, the paper proposes that finite-width operators can be studied by considering their scaling behaviour. Some finite-width corrections will be irrelevant and disappear when the network becomes wider or if the effective theory is coarse-grained. But others may be relevant or marginal, meaning that they can be important to understand and shape the distribution of function space and might lead to critical regimes. The term criticality is used here in a very specific sense. It is not just a description of neural networks as a combination of a weak similarity of a network with a physical system, it is not just the repetition of the common claim that neural networks exist in the vicinity of chaos. Criticality is the state where a system becomes sensitive to scale, correlation structure, and perturbation. In statistical physics critical points are associated with long-range correlations, scaling behaviour, universality and control over operators. In neural network function space, criticality can be formulated in terms of behaviour of correlation functions and effective couplings. A neural network ensemble can be considered as being near a critical regime when higher order connected correlation functions become non-negligible, when the induced function distribution becomes sensitive to changes in control parameters and when finite width interaction terms get relevant or marginal scaling behaviour. This is not defined by parameter count alone or training loss alone but by the structure of the probability distribution over functions \cite{wilson1974,cardy1996,schoenholz2017,poole2016}.

We also reframe the concept of overparameterization. In a normal parameter-space language, overparameterization seems to increase complexity since the number of parameters to train becomes very large. However, in the neural network–quantum field theory correspondence, increasing the width may lead the function space distribution closer to the Gaussian process limit. From the field theoretical point of view, increasing the width may reduce the interaction and make the theory nearer to a free-field fixed point. Overparameterization may come with higher functional complexity, as opposed to the functional simplicity as it is now, at least in the prior or weakly trained regime. A network can contain many parameters but still produce a fairly simple Gaussian distribution over function. Conversely, a network with a finite width or trained network may have fewer parameters but behave more interactively on function space as the higher order connected correlations are no longer negligible. This difference between parameter space complexity and function space complexity is one of the most important reasons for this work.

It is also possible to connect finite width corrections with expressivity. A purely Gaussian function space theory is completely characterized by its covariance kernel. While this is indeed convenient, it may not capture the full richness of finite neural networks. Non-Gaussian interactions introduce higher-order statistical structure, and this may be of importance for representing functions not well described by second order correlations alone. In this sense, finite width corrections may provide not only a break with the Gaussian process approximation, but might also be the field-theoretic mechanism for finite neural networks to appear expressive beyond the free theory. The same corrections may also lead to instability if they are too strong or if they move the system into a poorly controlled region of function space. Criticality can thus be understood as a boundary phenomenon: the regime in which finite-width interactions are strong enough to generate expressive structure, but not so uncontrolled that the distribution in function space is unstable or untrainable \cite{halverson2021,grosvenor2022}.

Training is a further layer to this picture. A randomly initialized neural network defines a prior distribution over functions, but learning changes its distribution. The network ensemble is formed by data, loss, learning rate, regularization, and architecture. In function space, training can be thought of as a deformation of effective action. This does not mean that gradient descent is the same as Wilsonian renormalization. The two processes have different definitions and operate in different spaces. But both can be understood as flows that change the effective description of a system. Training moves the network through parameter space and it also changes the induced distribution over functions. The Wilsonian perspective suggests that the important question is not where the parameters are moving, but how the effective function space couplings evolve. The generalization can then be related to whether training suppresses nonuniversal or highly data-specific interactions while maintaining relevant structures for prediction. Overfitting may correspond to the amplification of interactions that fit the training data but don't define stable or universal function space behaviour \cite{jacot2018,halverson2021}.

In this paper, we propose a structured approach for interpreting finite width corrections and criticality in the neural network - quantum field theory correspondence. The Gaussian-process limit is regarded as a free-field fixed point. Finite width corrections are treated as interactions. Criticality is defined in terms of scaling correlations and effective couplings. Overparameterization is presented as a process of getting closer to functional simplicity. Trainability and generalization are connected to the position of the network ensemble relative to critical surfaces in function space. These assertions lay the foundation for more precise mathematical and empirical diagnostics of criticality in neural networks.

The paper is organized as follows. In Section 2, we introduce a framework to describe neural networks as probability distributions over functions, interpret the infinite width Gaussian process as a free field fixed point and describe the finite width corrections as non-Gaussian interaction terms in Wilsonian function space terms. We present the main results in Section 3, including kernel convergence, finite width cumulants, Gaussianity diagnostics, critical correlation propagation, coupling flow diagrams, width suppressed interactions and schematic phase structure. In Section 4, we discuss the importance and limitations of the framework for overparameterization, trainability, generalization, edge of chaos and the generalization in relation to the architecture as well as the potential for future testing on the theory. And finally, in Section 5 we discuss how the Wilsonian interpretation can be used to develop a more general study of finite width neural networks, criticality and interacting function space theories.

\section{Theoretical Framework}

A neural network can be understood as not only a computational model with parameters, but also as a probability distribution over functions. In the usual parameter space description in which we describe a neural network, a network consists of a set of weights and biases, which together with the architecture, activation function, loss function, and optimization procedure. For a fixed parameter vector, the network defines a deterministic map from input to output. However, if the parameters are randomly chosen, the network is no longer able to represent a single function. Instead, it is a network of possible functions that are generated by a different draw from the initial distribution. This provides a basis for a function-space description of neural networks in which the main object is not the parameter vector itself but the induced probability distribution over input-output maps. This shift is important because the behaviour of a neural network is ultimately determined by the function realized and not directly by the position of the parameter vector.

Let a neural network of width $N$ be denoted by $f_{\theta,N}: \mathcal{X} \rightarrow \mathbb{R}$, where $\theta$ are the trainable parameters and $\mathcal{X}$ is the input domain. If the parameters are drawn from a probability distribution $P(\theta)$, then our network induces a probability measure over functions, $P_N[f] = \int d\theta \, P(\theta) \,\delta[f-f_{\theta,N}]$. This formalizes the idea that the network architecture and the initialization define an ensemble of functions. The statistical structure of this ensemble can be studied by its correlation functions, $G_N^{(n)}(x_1,\ldots,x_n) = \left\langle f_{\theta,N}(x_1) \cdots f_{\theta,N}(x_n) \right\rangle_{\theta}$. Where the expectation is taken over the initialization distribution. The two-point function $G_N^{(2)}(x,x') = \left\langle f_{\theta,N}(x) f_{\theta,N}(x') \right\rangle_{\theta}$ plays the role of a covariance kernel. It measures how output at two input points are statistically related over the ensemble of randomly initialized networks. More complex correlations are introduced in higher order correlation functions as they encode more complex statistical dependency between outputs at different input locations.

For the infinite width limit, many neural network architectures converge to Gaussian processes. In such a limit, the function space distribution is Gaussian in every case as the two-point function, which means that all higher order correlation functions factorize by Wick's theorem. For instance, the four-point function in the Gaussian limit satisfies that $G^{(4)}(x_1,x_2,x_3,x_4) = G^{(2)}(x_1,x_2)G^{(2)}(x_3,x_4) + G^{(2)}(x_1,x_3)G^{(2)}(x_2,x_4) + G^{(2)}(x_1,x_4) G^{(2)}(x_2,x_3)$. The connected four-point function therefore vanishes:
\begin{equation*}
G^{(4)}_{c}(x_1,x_2,x_3,x_4)=0.
\end{equation*}
This vanishing of the connected higher-order correlations is the property of a Gaussian theory. It shows that at an infinite width, the network function distribution of neural networks does not have any independent higher-order statistical structure beyond the covariance kernel. The network may have infinitely many parameters in the network but its function space description is quite simple.

The neural network-quantum field theory correspondence is a natural way to explain this structure. In this correspondence, the network output $f(x)$ is treated as a field over $\mathcal{X}$. The function-space probability distribution can be expressed formally as an effective action:
\begin{equation}
P_N[f] = \frac{1}{Z_N}e^{-S_N[f]},
\end{equation}
where $Z_N = \int \mathcal{D}f\,e^{-S_N[f]}$ is the corresponding normalization functional. In the infinite-width limit, the effective action is Gaussian:
\begin{equation}
S_{\infty}[f] = \frac{1}{2} \int_{\mathcal{X}} dx \int_{\mathcal{X}} dx'\, f(x)K^{-1}(x,x')f(x'),
\end{equation}
where $K(x,x')$ is the kernel of the Gaussian process. This expression is exactly like the action of a free Euclidean field theory. The kernel $K(x,x')$ propagates the theory,
\begin{equation}
\left\langle f(x)f(x') \right\rangle = K(x, x').
\end{equation}
The infinite-width neural network is therefore analogous to a free field theory because its effective action is quadratic and the correlation functions are determined by the two-point function \cite{halverson2021,grosvenor2022}.

The importance of this correspondence becomes clear when we consider finite-width effects. A finite-width neural network does not always produce a Gaussian distribution over functions. Its higher-order connected correlation functions do not vanish exactly, and its action cannot always be written in quadratic form. At finite width, the action can be written as
\begin{equation}
S_N[f] = S_{\infty}[f] + \Delta S_N[f],
\end{equation}
where $\Delta S_N[f]$ represents finite-width corrections. The corrections can be expanded in terms of non-Gaussian operators:
\begin{equation}
\Delta S_N[f] = \sum_{k \geq 3} \frac{1}{k!} \int dx_1 \cdots dx_k\, \lambda_k^{(N)}(x_1,\ldots,x_k) f(x_1) \cdots f(x_k).
\end{equation}
The functions $\lambda_k^{(N)}$ represent finite-width effective couplings, and they encode the strength and structure of the non-Gaussian interactions generated by finite network width. From the field-theoretic perspective, finite-width corrections are therefore not merely approximation errors around the infinite-width limit. They are interaction terms in an effective field theory of neural network functions \cite{halverson2021,erbin2021}.

This is the interpretation we will be working on in this paper. If the infinite-width Gaussian-process limit corresponds to a free theory, then finite-width neural networks correspond to interacting effective theories. The nonzero connected correlation functions of finite-width networks are the field-theoretic manifestation of these interactions. The connected four-point function is particularly important as it is the lowest-order diagnostic of non-Gaussianity in many symmetric cases:
\begin{equation}
G^{(4)}_{N,c}(x_1,x_2,x_3,x_4) = G^{(4)}_N(x_1,x_2,x_3,x_4) - G^{(4)}_{\mathrm{Wick}}(x_1,x_2,x_3,x_4).
\end{equation}
Here $G^{(4)}_{\mathrm{Wick}}$ denotes the Gaussian factorization into products of two-point functions. If this connected term is negligible, the network remains close to the Gaussian-process description. If it becomes significant, the finite-width network displays interaction structure that cannot be captured by the kernel alone.

The finite-width expansion can be expressed schematically as
\begin{equation}
G_N^{(n)} = G_{\infty}^{(n)} + \frac{1}{N}G_1^{(n)} + \frac{1}{N^2}G_2^{(n)} + \cdots.
\end{equation}
For $n > 2$, the connected component generally satisfies a scaling relation of the form
\begin{equation}
G^{(n)}_{N,c} = \frac{1}{N^{\alpha_n}}C_n(x_1,\ldots,x_n) + \mathcal{O}(N^{-\alpha_n-1}),
\end{equation}
where $\alpha_n > 0$ and $C_n$ captures the leading finite-width contribution. This scaling relation makes width a field-theoretic control parameter. As $N$ increases, finite-width interactions are suppressed and the network approaches the Gaussian-process limit. As $N$ decreases, or as training and architecture amplify non-Gaussian structure, connected higher-order correlations become more important. The degree of departure from Gaussianity can therefore be interpreted as a measure of interaction strength in neural network function space.

This leads naturally to a Wilsonian interpretation of the neural network-quantum field theory correspondence. In Wilsonian field theory, a fixed point is a theory whose structure is invariant under changes of scale. Perturbations around a fixed point are classified according to their behaviour under renormalization-group flow. Relevant perturbations grow under coarse-graining and determine large-scale behaviour. Irrelevant perturbations decay and become less important. Marginal perturbations remain at the boundary and may control critical or near-critical phenomena. The proposal developed here is that the infinite-width Gaussian-process limit can be treated as a Wilsonian fixed point in neural network function space, while finite-width corrections act as perturbations around that fixed point \cite{wilson1974,cardy1996,zinnjustin2002,erbin2021}.

Let the effective action be written as
\begin{equation}
S[f] = S_{*}[f] + \sum_i g_i\mathcal{O}_i[f],
\end{equation}
where $S_{*}[f]$ denotes the Gaussian fixed-point action, $\mathcal{O}_i[f]$ are perturbing operators, and $g_i$ are the associated effective couplings. A renormalization-group transformation induces a flow in the space of couplings,
\begin{equation}
\frac{dg_i}{d\ell} = \beta_i(\{g\}),
\end{equation}
where $\ell$ is an RG scale parameter and $\beta_i$ is the beta function associated with $g_i$. Near the fixed point, this flow may be linearized:
\begin{equation}
\frac{dg_i}{d\ell} = y_i g_i + \mathcal{O}(g^2),
\end{equation}
where $y_i$ is the scaling exponent of the perturbation. If $y_i > 0$, the perturbation is relevant. If $y_i < 0$, it is irrelevant. If $y_i = 0$, it is marginal. The classification of finite-width operators according to this Wilsonian scheme provides a way to distinguish between finite-width effects that are merely small technical corrections and finite-width effects that can qualitatively alter the behaviour of the neural network ensemble.

Criticality in neural network function space can then be defined through the scaling behaviour of these finite-width perturbations. A neural network ensemble may be considered near a critical regime when higher-order connected correlation functions become significant, when one or more finite-width operators become relevant or marginal, and when the induced distribution over functions becomes sensitive to control parameters such as width, depth, activation nonlinearity, initialization variance, learning rate, or training time. A critical point or critical surface may be represented by a condition of the form
\begin{equation}
\beta_i(g_c) = 0
\end{equation}
for one or more effective couplings. More generally, if the network depends on a collection of control parameters
\begin{equation}
\bm{\alpha} = (N,L,\sigma_w,\sigma_b,\phi,\eta,t),
\end{equation}
where $N$ is width, $L$ is depth, $\sigma_w$ is weight variance, $\sigma_b$ is bias variance, $\phi$ is the activation function, $\eta$ is the learning rate, and $t$ is training time, then the effective couplings may be written as
\begin{equation}
g_i = g_i(\bm{\alpha}).
\end{equation}
A critical surface can therefore be described schematically as
\begin{equation}
\mathcal{C} = \left\{ \bm{\alpha}: \beta_i(g(\bm{\alpha})) \approx 0 \text{ for one or more relevant or marginal couplings} \right\}.
\end{equation}
Criticality is not defined simply by a large number of parameters, low loss, or empirical success. It is defined by the behaviour of the induced function-space theory under changes of scale and control parameters \cite{wilson1974,cardy1996,erbin2021,erbin2022}.

This model can also reframe the image of the edge of chaos in deep learning theory. The edge of chaos is a boundary between ordered and chaotic signal propagation. In an ordered regime, correlations between inputs may break as signals propagate through depth, and representations may become less distinguishable and gradients become less and less detectable. In a chaotic regime, large differences in inputs or parameters may be amplified and the signal propagates with extreme instability and exploding gradients. Near the boundary between these two regimes, signal propagation may be balanced over many layers, making it trainable. The Wilsonian function space interpretation does not replace this picture, but it provides it with a broader theoretical setting. The edge of chaos can be viewed as an important surface in the effective field theory of neural network functions of finite-width interactions and correlation propagation. In this view, the important object is not just the layerwise propagation of correlations to the network but also the whole scaling structure of the induced function distribution \cite{schoenholz2017,poole2016,grosvenor2022}.

The same perspective explains the relationship between overparameterization and functional simplicity. In parameter space, overparameterization seems to increase complexity because the number of trainable degrees of freedom is very large. In function space, however, increasing width might suppress the finite-width interaction terms and move the network closer to the Gaussian fixed point: \(N \rightarrow \infty\). \(g_i(N) \rightarrow 0\) for finite-width interaction couplings. Equivalently, \(S_N[f] \rightarrow S_{\infty}[f]\) as \(N \rightarrow \infty\). Thus, overparameterization may be a functional simplification rather than a functional complexity, at least in the initialization regime and weakly interacting approximations. A highly overparameterized network could contain many parameters but still have a relatively simple Gaussian distribution over functions. On the other hand, a finite-width or highly trained network may have a richer and more complex effective action since its higher-order correlations are no longer negligible. This distinction between parameter space complexity and function space complexity is essential to understand why a large neural network can generalize well even with a lot more parameters than data points.

Finite-width corrections also provide a field-theoretic route to expressivity. The purely Gaussian function space theory is completely determined by its covariance kernel. This is powerful, but not without its drawbacks because no higher-order statistical structure is present in the theory. Finite-width interactions can lead to non-Gaussian dependence among outputs at different input points. Such interactions might lead to a network ensemble that has richer structures than the ones present in the free theory. From this perspective, expressivity does not come from the number of parameters or the depth of the distribution of functions; it comes from the interaction structure of the distribution of functions. The finite-width effects that enhance expressivity can also lead to instability if they get too strong or poorly controlled. Criticality is then a boundary condition where finite-width interactions are strong enough to construct expressive structure, but not enough to allow for uncontrolled instability \cite{halverson2021,grosvenor2022}.

Training complicates and enriches this picture even more. A randomly initialized network will have a distribution of functions, but training changes it. The network ensemble is formed by data, loss, optimization dynamics, learning rate, regularization, and architecture. In function space terms, training changes the effective action: \(S_{\mathrm{init}}[f] \rightarrow S_{\mathrm{trained}}[f]\). In figure, this trained action is shown schematically as \(S_{\mathrm{trained}}[f] = S_{\infty}[f] + \sum_i g_i(t) \mathcal{O}_i[f]\), where the effective couplings depend on training time. This does not imply that gradient descent is the same as Wilsonian renormalization. Gradient descent is an optimization flow in parameter space and Wilsonian renormalization is a scale transformation in the space of effective theories. However, both can be understood as flows that change the description of a system. Training not only changes the location of the parameter vector but also the induced distribution over functions. The Wilsonian question is how training changes the effective couplings, whether it stops irrelevant nonuniversal interactions, and whether it drives the network towards or away from critical surfaces \cite{jacot2018,halverson2021}.

This view suggests a possible field-theoretic interpretation of generalization. A well-generalizing network is one in which the training dynamics can preserve or improve relevant structures while keeping the mostly data-specific irrelevant interactions out of the picture. An overfitting network might further strengthen nonuniversal couplings that capture idiosyncratic details of the training set but do not describe stable function space behaviour beyond it. Generalization might depend not only on explicit regularization or number of parameters but also on the shape of the effective action after training. In this way, implicit regularization is understood as an optimization to favor those trajectories which are near simple, weakly interacting, or critically balanced regions of function space.

The Wilsonian framework also offers another way to think about architectural universality. In statistical physics, systems with different microscopic details can have the same large-scale critical behaviour if they belong to the same universality class. A similar idea can be applied to neural network architectures. Different architectures may have different parameterizations, but they may share similar Gaussian-process limits, symmetry structures, or relevant finite-width perturbations. A fully connected network, a convolutional network, a recurrent network, a graph neural network, or a transformer-like architecture may differ substantially in implementation, yet still have some function space fixed points or scaling behaviours. In this setting, an architecture can be characterized by its Gaussian kernel and its symmetries in its induced function distributions and the finite-width operators that perturb it away from the Gaussian fixed point. Two architectures can be considered to be part of the same effective universality class if they move towards the same fixed point and have the same perturbations \cite{cardy1996,halverson2021}.

This architectural interpretation changes the way neural networks are compared. Instead of comparing architectures only on the basis of empirical quality, number of parameters, or computational cost, we can examine the Gaussian fixed point of each architecture, the symmetries of its effective action, the finite-width operators that can be allowed, the relevance or marginality of these operators, how fast the architecture approaches the Gaussian limit with increasing width, and the deformation of effective couplings during training. These considerations point to a classification of neural networks on the basis of their effective field-theoretic behaviour, not on the basis of engineering design alone. The proposed framework also provides diagnostic tools for measuring criticality. The first diagnostic is the magnitude of connected higher-order correlation functions. The connected four-point function, \( G^{(4)}_c = G^{(4)} - G^{(4)}_{\mathrm{Wick}} \), is the simplest nontrivial measure of non-Gaussianity. If this term is small, the network remains close to the Gaussian fixed point. If it becomes significant, finite-width interactions are active. The second diagnostic is the scaling of connected correlations with width, \( G^{(n)}_{N,c} \sim N^{-\alpha_n} \). The exponents \( \alpha_n \) indicate how rapidly the theory approaches the Gaussian limit. Changes in these exponents across architectures, activation functions, or training regimes may indicate different universality classes or different forms of finite-width interaction. The third diagnostic is the flow of effective couplings under changes in control parameters. For instance, varying the initialization variance may induce a trajectory in coupling space: \( \sigma_w \mapsto g_i(\sigma_w) \). If one or more couplings approach marginality at a particular value of \( \sigma_w \), this may indicate a critical initialization regime \cite{halverson2021,erbin2021,grosvenor2022}.

A fourth diagnostic is correlation length. In deep neural networks, the propagation of correlations through layers can be associated with a characteristic depth scale. If it is short, the correlations decay or become unstable very quickly. If it is large, the network may be able to retain structured information over depth and operate near a critical regime. In field theory, if the correlation length is large, that is the network is sensitive to scale and is likely to behave at a universal level. The correlation length may thus provide a link between the edge-of-chaos literature and field theory in function space. It relates layerwise signal propagation with the question of whether the neural network ensemble is located near a critical surface in effective coupling space \cite{schoenholz2017,poole2016,grosvenor2022}.

The framework we have developed is purely conceptual, but it is not just metaphorical. It identifies mathematical objects that are easily obtainable to us: the function-space probability distribution, the effective action, the Gaussian kernel, connected higher-order correlation functions, finite-width couplings, beta functions, scaling exponents and critical surfaces. Such objects open up the way to a more systematic theory of neural network criticality. The critical behaviour of neural networks should not only be understood via optimisation dynamics or parameter space geometry (the same for the mathematical description, but also the scaling structure of the induced distribution over functions) but also the network quantum field theory. It is a way to understand how free Gaussian behaviour, finite-width interaction, expressivity, trainability, overparameterization, and architectural universality are related in a single effective theory \cite{halverson2021,grosvenor2022,erbin2021,erbin2022}.

\newcommand{\alignedsubfig}[3]{%
\begin{subfigure}[t]{0.485\textwidth}
\centering
\vspace{0pt}
\includegraphics[height=5.1cm,width=\linewidth,keepaspectratio]{#1}
\caption{#2}
\label{#3}
\end{subfigure}
}

\section{Results}

The Wilsonian interpretation developed in this paper is best realized when we actually address our abstract objects of the neural network -quantum field theory correspondence with the objective of analyzing them with tools in the form of function space diagnostics. In this regard, the relevant quantities are not classification accuracy, training loss, or benchmark performance, but the mathematical characteristics of a neural network ensemble as it edges and breaks out of the Gaussian process. The empirical kernel, the finite width cumulants, the output distribution, the correlation propagation scale, and the schematically drawn effective couplings are all related to a single point of view of the theory.

The primary structure of the system as defined by these diagnostics is the separation of the Gaussian fixed point from the finite width departure from the Gaussian. We have, at infinite width, the corresponding distribution over functions is described by a Gaussian process and is characterized by the two point covariance kernel; this is a function-space analog to a free field theory. At finite width the network still has nonzero higher order connected correlations, and these can be viewed as interaction terms in an effective action. In this sense the criticality in our analysis is not at all related to the complexity and the number of parameters. It occurs when non-Gaussian finite width structure is controlled, scale sensitive and organized around a fixed point description. The plots below are therefore useful in theoretical diagnostics of the convergence of fixed points, their suppression, Gaussianization, critical correlation propagation and Wilsonian coupling.

A natural starting point is the finite width empirical kernel. For a random-feature ReLU network of the form
\begin{equation}
f_N(x)=\frac{1}{\sqrt{N}}\sum_{j=1}^{N}a_j\phi(w_j\cdot x),
\end{equation}
the empirical neural network Gaussian process kernel is given by
\begin{equation}
K_N(x,x') = \frac{1}{N}\sum_{j=1}^{N}\phi(w_j\cdot x)\phi(w_j\cdot x').
\end{equation}
The corresponding infinite width object is the limiting kernel $K_{\infty}(x,x')$, which is the propagator of the Gaussian function-space theory:
\begin{equation}
\langle f(x)f(x')\rangle = K_{\infty}(x,x').
\end{equation}
Since $K_N$ is an average over $N$ independent random features and is a standard average over $N$, its deviation from $K_{\infty}$ is expected to follow the central limit scaling
\begin{equation}
\frac{\|K_N-K_{\infty}\|_F}{\|K_{\infty}\|_F} \sim N^{-1/2}.
\end{equation}
This relation is useful in practice since it indicates that the width increases with the convergence towards the Gaussian fixed point propagator. Width therefore has two functions: it makes the parameterization bigger and reduces the finite correlations in the covariance structure in the induced function space.

\begin{figure}[H]
\centering
\alignedsubfig{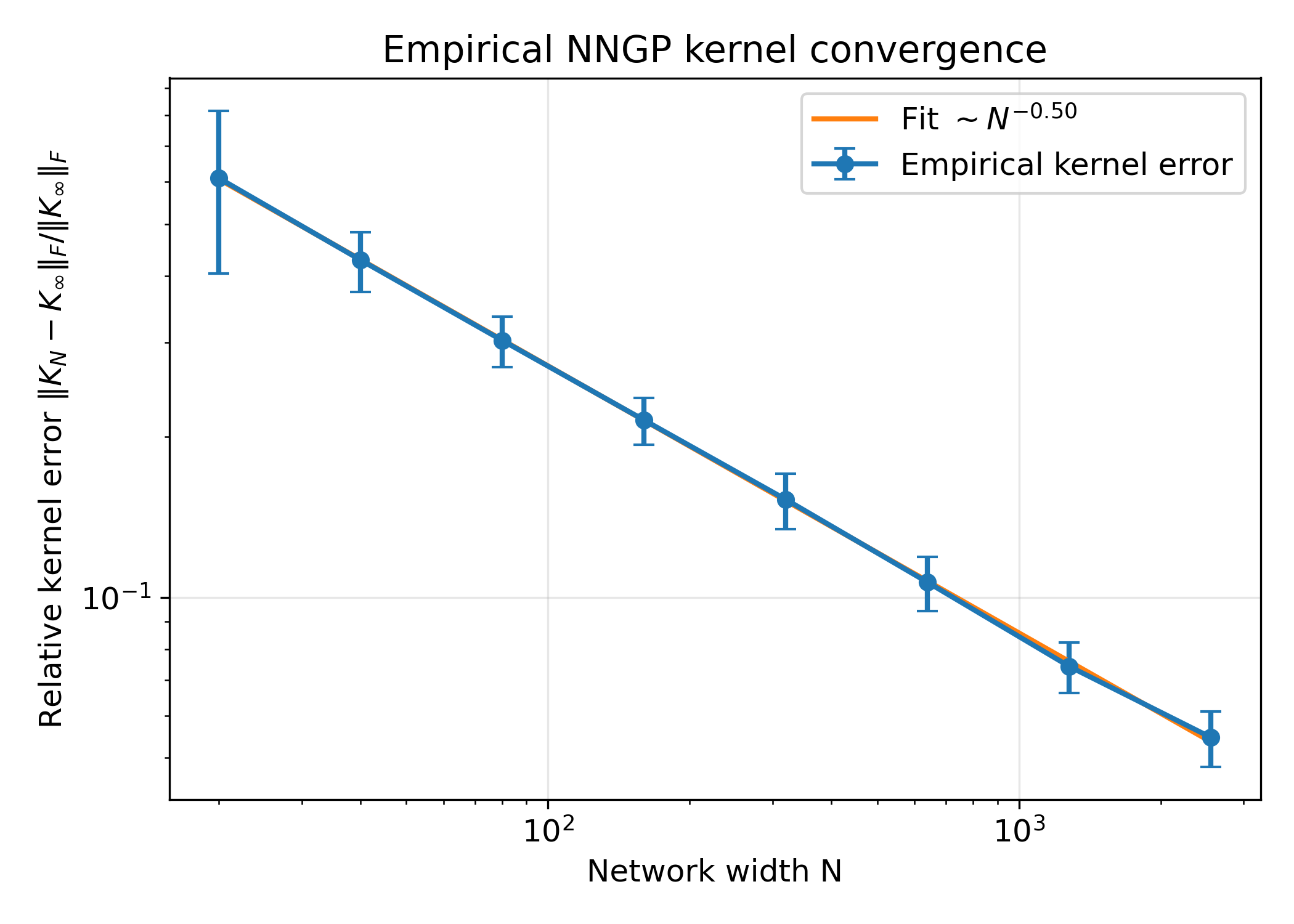}
{Relative convergence of the empirical NNGP kernel to the infinite-width kernel.}
{fig:kernel-convergence-panel}
\hfill
\alignedsubfig{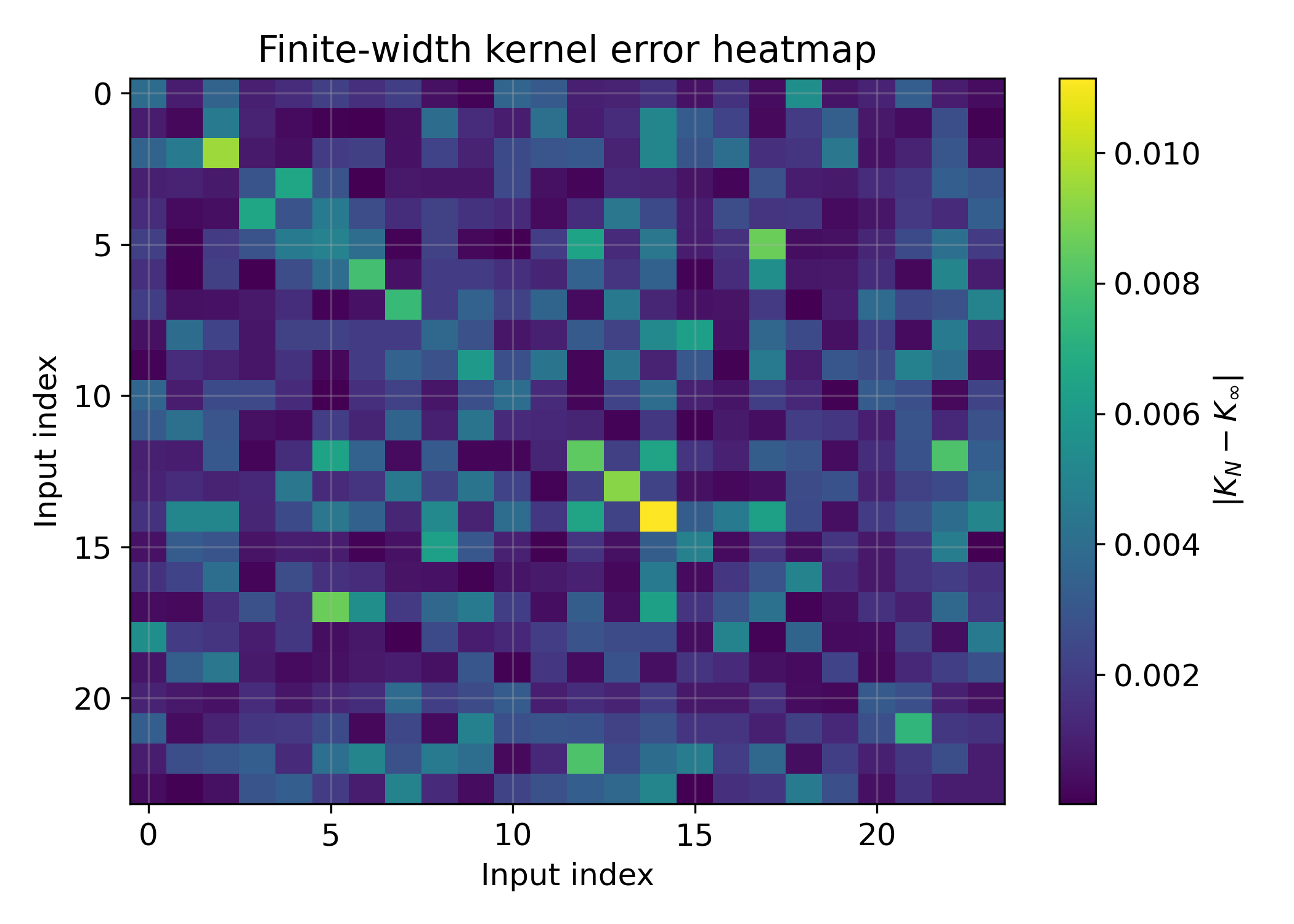}
{Finite-width kernel error heatmap for a typical empirical kernel.}
{fig:kernel-error-panel}
\caption{Kernel-level convergence toward the Gaussian-process fixed point. The kernel error is as expected at the central limit decay, and the heatmap shows that finite-width deviations are structured by input pairs.}
\label{fig:kernel-convergence-error}
\end{figure}

Figure~\ref{fig:kernel-convergence-error} shows the convergence of kernel-convergence-error. In the left panel we can see the relative Frobenius error between finite-width empirical kernel and analytic infinite-width kernel. The slope of the fit agrees with the expected $N^{-1/2}$ central limit decay. In the right panel we see the finite-width error as a heatmap, which clearly shows that the residual deviation from the limit kernel depends on the input pairs. This is important because finite-width effects are not only scalar perturbations, but also input-dependent deviations from the free propagator. In the language of effective field theory, this is the first visible indication that a finite network is close to, but not exactly at, the Gaussian fixed point.

\begin{figure}[H]
\centering
\alignedsubfig{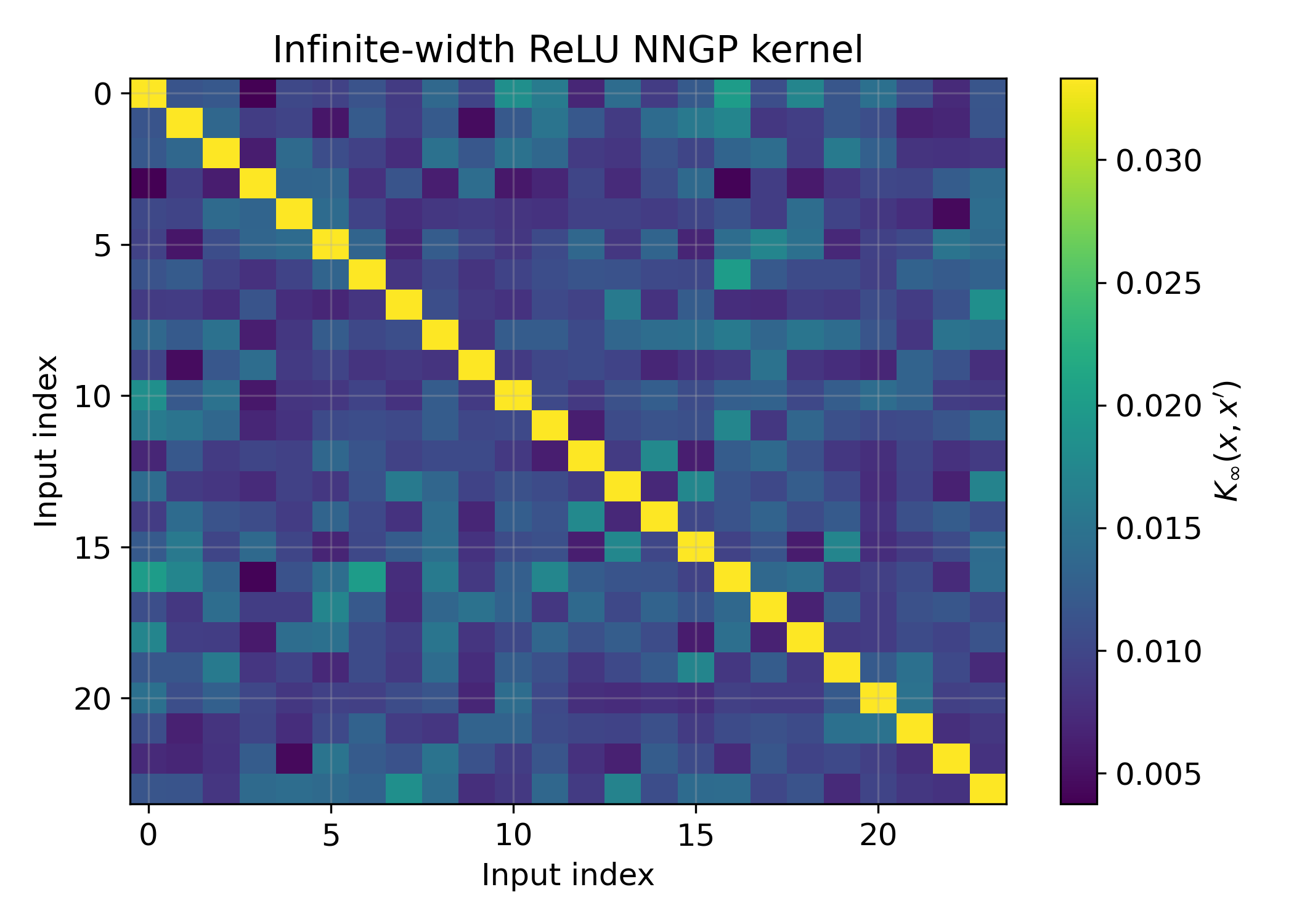}
{Infinite-width ReLU NNGP kernel $K_{\infty}(x,x')$.}
{fig:kernel-infinite-panel}
\hfill
\alignedsubfig{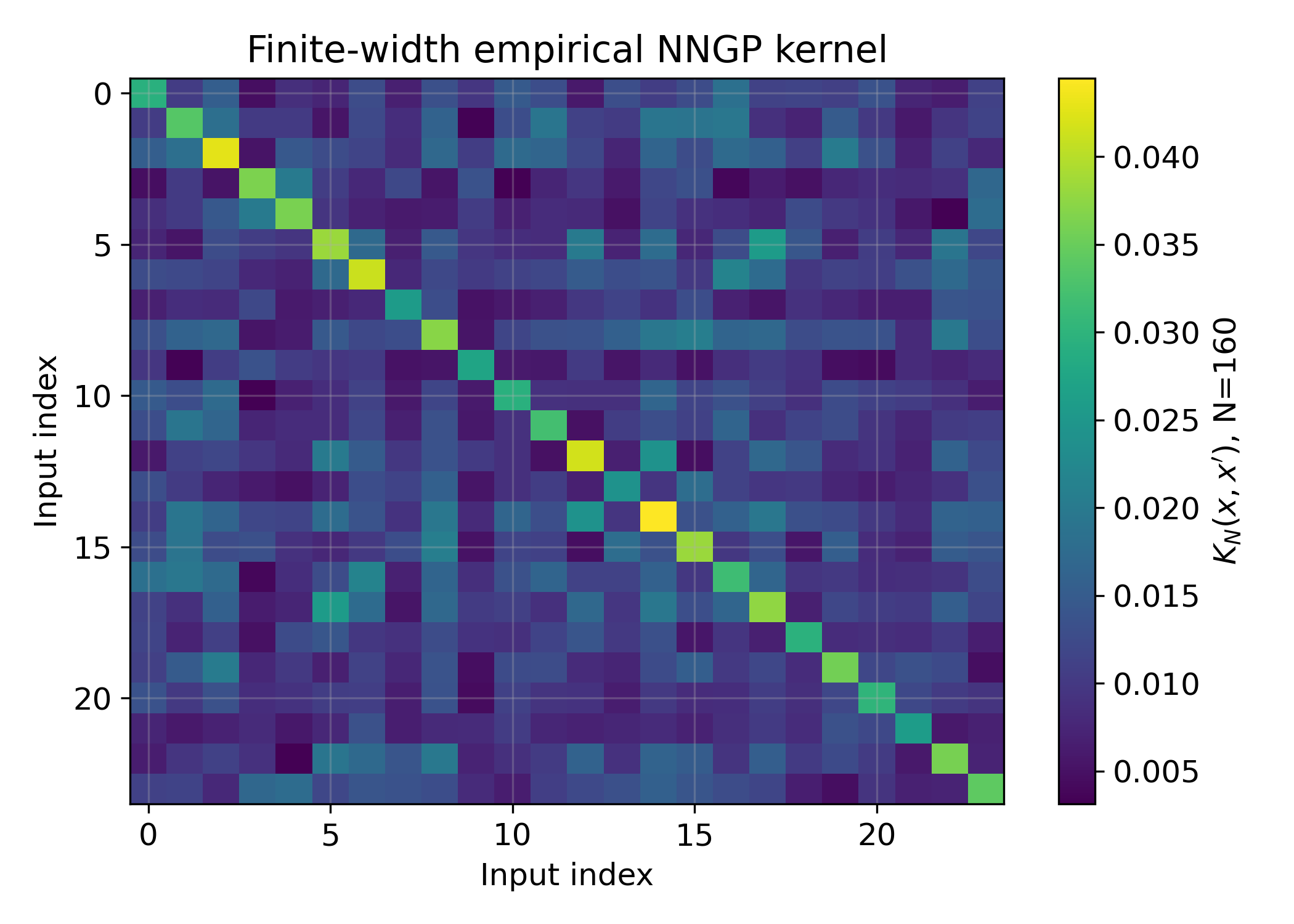}
{Finite-width empirical kernel $K_N(x,x')$.}
{fig:kernel-finite-panel}
\caption{Comparison between the analytic infinite-width kernel and a finite-width empirical kernel. The similarity of the covariance structure provides the basis for the interpretation of the Gaussian process limit as the free-field reference theory of the neural network ensemble.}
\label{fig:kernel-heatmaps}
\end{figure}

The covariance structure itself is shown in Fig.~\ref{fig:kernel-heatmaps}. The left panel gives the analytic infinite-width ReLU NNGP kernel, while the right panel shows a finite-width empirical kernel generated from the same set of input. The finite-width kernel gives the broad shape of the limiting covariance matrix, but it retains the expected fluctuations of the empirical kernel. This comparison of the two-point kernel is important because the two-point kernel is the real object of the Gaussian process limit in the field of theory. The close agreement between $K_N$ and $K_{\infty}$ indicates that the finite network approaches the covariance structure of the infinite width fixed point.

This helps us distinguish between parameter space size and function space complexity. A network might become wider and thus contain more parameters, but the distribution of the parameters on the function is constrained by a Gaussian fixed point kernel. Increasing the parameter space of a network also does not necessarily imply that the interaction between the parameters in the network is more complicated. On the other hand, if the number of features is increasing in the previous random features context, the distribution of the parameters on the function is more effectively based on a simple free theory. This is one of the main implications of considering neural networks in function space rather than in parameter space.

Kernel convergence gives us the behaviour of the two-point function, but the difference between free and interacting function space theories is the higher order connected correlations. A Gaussian process is entirely determined by its covariance kernel, and all connected correlation functions beyond second order vanish:
\begin{equation}
G^{(n)}_{\infty,c}(x_1,\ldots,x_n)=0,
\qquad n>2.
\end{equation}
At finite width, this Gaussian factorization is no longer exact. The effective action yields non-Gaussian terms,
\begin{equation}
S_N[f] = S_{\infty}[f] + \sum_{k\geq 3} \frac{1}{k!} \int dx_1\cdots dx_k\, \lambda_k^{(N)}(x_1,\ldots,x_k) f(x_1)\cdots f(x_k),
\end{equation}
where $\lambda_k^{(N)}$ may be interpreted as finite width effective couplings. These terms are the function space analogues of interaction vertices in an effective field theory. Their presence means that a finite width neural network is not only an approximate Gaussian process; it is an interacting theory whose departure from the Gaussian fixed point is encoded in its higher order structure \cite{halverson2021,erbin2021}.

The lowest-order diagnostic of this interaction structure is the connected four-point function, 
\begin{equation}
G^{(4)}_{N,c}
=
G^{(4)}_N
-
G^{(4)}_{\mathrm{Wick}},
\end{equation}
where $G^{(4)}_{\mathrm{Wick}}$ is the Gaussian Wick factorization into products of the two-point functions. In this random feature model we now consider, the scale of this finite-width correction is generated by the cumulant structure of the normalized sums. If 
\begin{equation}
f_N(x)=\frac{1}{\sqrt{N}}\sum_{j=1}^{N}z_j(x),
\end{equation}
where $z_j$ are independent feature contributions, then the fourth cumulant follows 
\begin{equation}
\kappa_4[f_N]
=
\frac{1}{N}\kappa_4[z].
\end{equation}
Similarly, the one-point excess kurtosis scales as 
\begin{equation}
\gamma_2[f_N]
=
\frac{1}{N}\gamma_2[z].
\end{equation}
The scaling relations of the above results provide a clean and analytically justified representation of finite-width non-Gaussianity \cite{halverson2021,grosvenor2022}.

This finite-width scaling is important since it gives the non-Gaussian correction a precise status. We do not consider the connected cumulants to be accidental numerical deviations at all, and we do not consider them as defects of the infinite-width approximation, but as the measurable residues of the interaction in the network ensemble. Just as an interacting field theory is distinct from a free field theory if there are connected higher-order correlation functions, a finite-width neural network is distinct from a Gaussian-process network when there are nonzero cumulants. The fact that these cumulants decrease with width supports the view that the Gaussian-process limit is an organizing fixed point and the finite-width networks are contained in a controlled neighborhood around this point.

The interpretation also clarifies why finite-width effects may not be dismissed entirely in the short term. Although the cumulant corrections disappear as $N \rightarrow \infty$, in practice neural networks are always finite and the expressive behavior of these networks may depend on precisely the terms absent in the Gaussian theory. A purely Gaussian process is, in fact, completely specified by its covariance kernel, whereas a finite network can have higher-order statistical structure which cannot be recovered from the kernel. This additional structure may affect the model's representation of data, and the sensitivity to training, and can be used to get away from a purely kernel-like model. In this way finite-width corrections are not just restrictions of the approximation, but they are how the finite network becomes an interacting function space model.

This is a good comparison between convergence and expressivity. Kernel convergence demonstrates that the two-point structure of the finite network approaches the infinite-width propagator, but the suppression of higher-order cumulants shows that there is still interaction beyond that propagator. A network can therefore be close to the Gaussian fixed point at the level of its covariance, but still have a finite-width non-Gaussian structure at higher orders. This is critical in the Wilsonian interpretation, because the effective theory is not only about the fixed point itself but also the perturbations around it. The question therefore is not whether the network is close to the Gaussian process, but which finite-width perturbations are still present and how they increase or decrease with the width, depth, initialization, and training of the network.

\begin{figure}[H]
\centering
\alignedsubfig{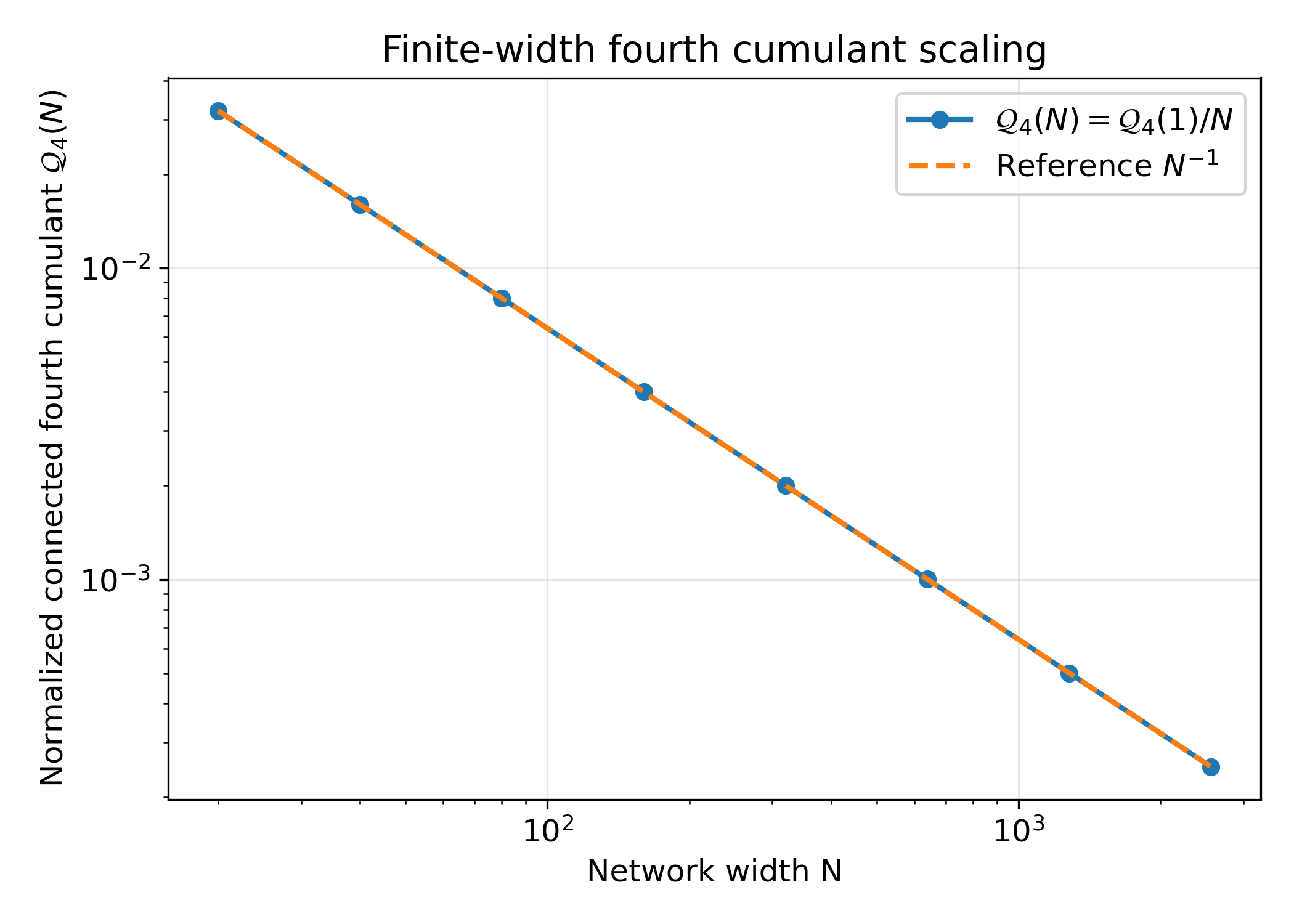}
{Connected fourth cumulant scaling as $1/N$.}
{fig:fourth-cumulant-panel}
\hfill
\alignedsubfig{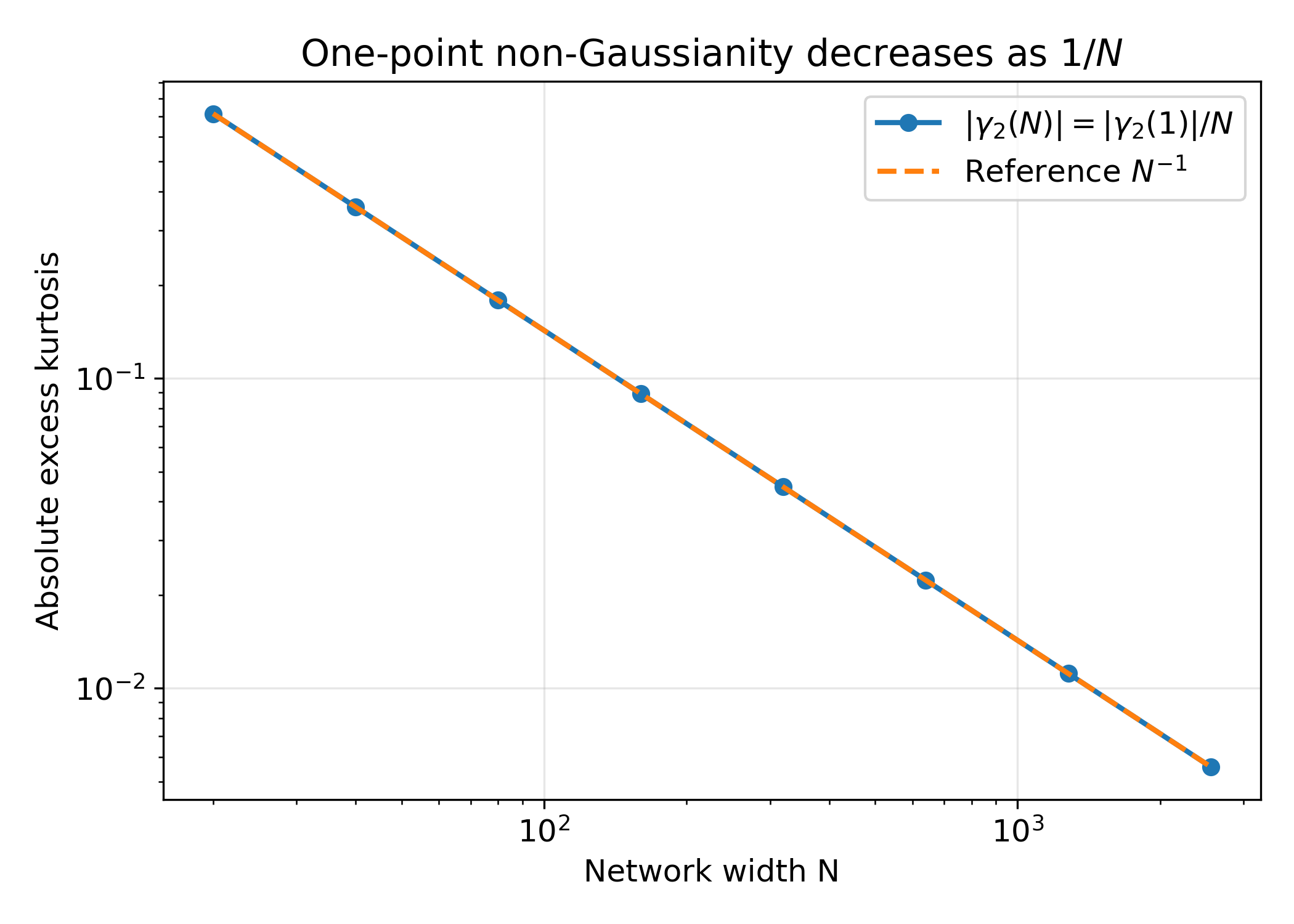}
{One-point excess kurtosis scaling as $1/N$.}
{fig:kurtosis-panel}
\caption{Finite-width non-Gaussianity as an interaction correction. The fourth cumulant and excess kurtosis both decrease as $1/N$, supporting the interpretation of finite-width effects as perturbative non-Gaussian corrections around the Gaussian fixed point.}
\label{fig:cumulant-kurtosis}
\end{figure}

Figure~\ref{fig:cumulant-kurtosis} shows the resulting finite-width suppression. The left panel shows the normalized fourth cumulant and the right one the absolute excess kurtosis. Both decrease as $1/N$, which is precisely the expected scaling for cumulants of normalized sums of independent feature contributions. This supports the notion that finite-width corrections are perturbative interaction terms around the Gaussian fixed point. The corrections become weaker with width increasing but at finite width they are still the mathematical source of the non-Gaussian structure. From the Wilsonian point of view, these quantities are not incidental errors; they are the finite-width operators whose scaling determines how the network deviates from the free theory.

The cumulant approach above also suggests a natural hierarchy of function-space non-Gaussianity diagnostics. A normalized fourth-order measure can be written as 
\begin{equation}
\mathcal{Q}_4(N)
=
\frac{
\|G^{(4)}_{N,c}\|
}{
\|G^{(4)}_{\mathrm{Wick}}\|
}.
\end{equation}
For a Gaussian process, this quantity vanishes as $N \rightarrow \infty$:
\begin{equation}
\lim_{N\rightarrow\infty}\mathcal{Q}_4(N)=0.
\end{equation}
More generally, we can define 
\begin{equation}
\mathcal{Q}_n(N)
=
\frac{
\|G^{(n)}_{N,c}\|
}{
\|G^{(n)}_{\mathrm{Gaussian}}\|
},
\qquad n>2.
\end{equation}
This hierarchy separates function-space interaction strength from the number of parameters. A network may be highly over-parameterized but close to a Gaussian fixed point if the higher-order functions are suppressed. A smaller or strongly trained network would have fewer parameters but be more connected. Function-space complexity is therefore controlled not only by the number of parameters but also by the magnitude and scaling of non-Gaussian correlations.

\begin{figure}[H]
\centering
\alignedsubfig{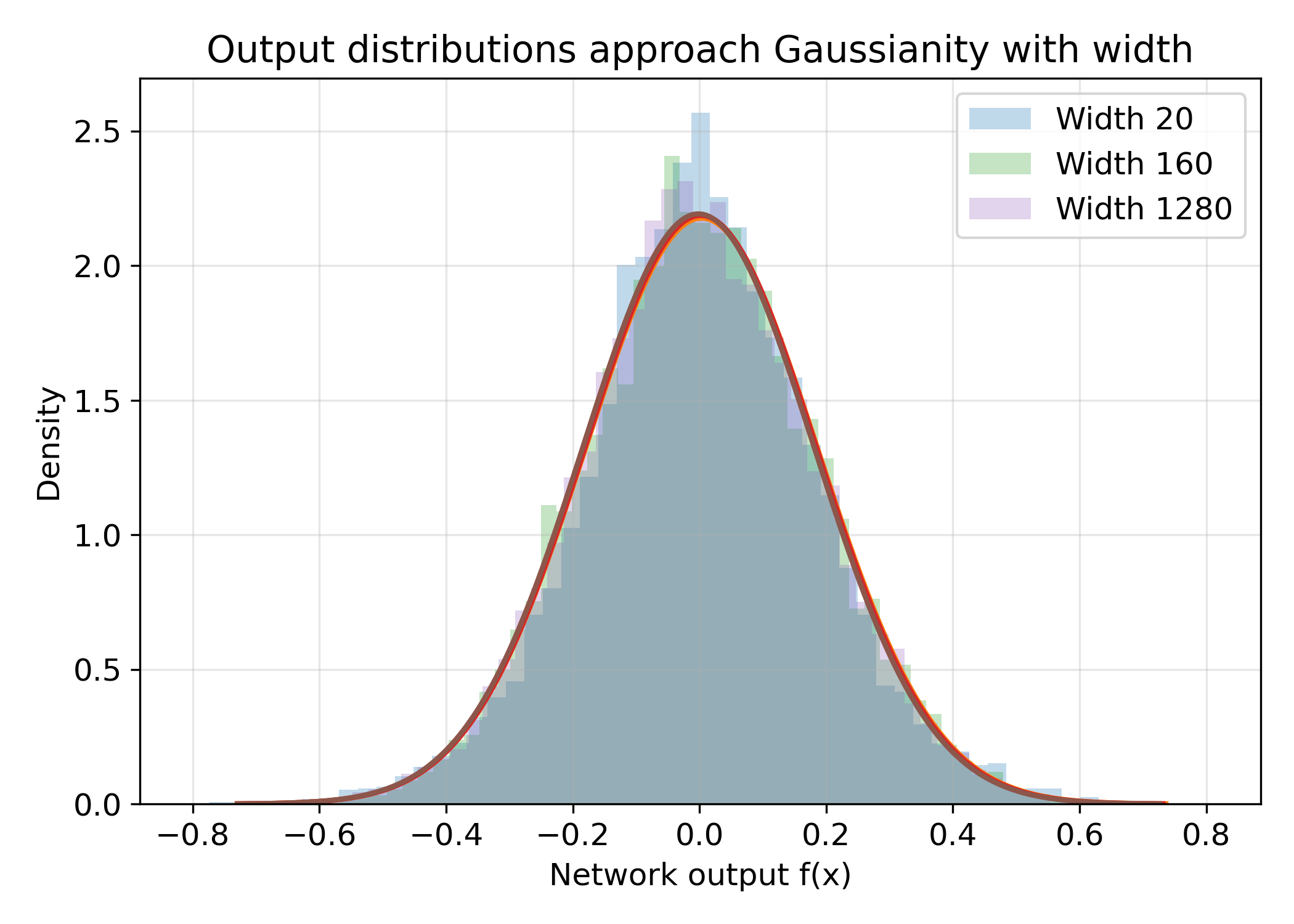}
{Output distributions for increasing network width.}
{fig:output-distribution-panel}
\hfill
\alignedsubfig{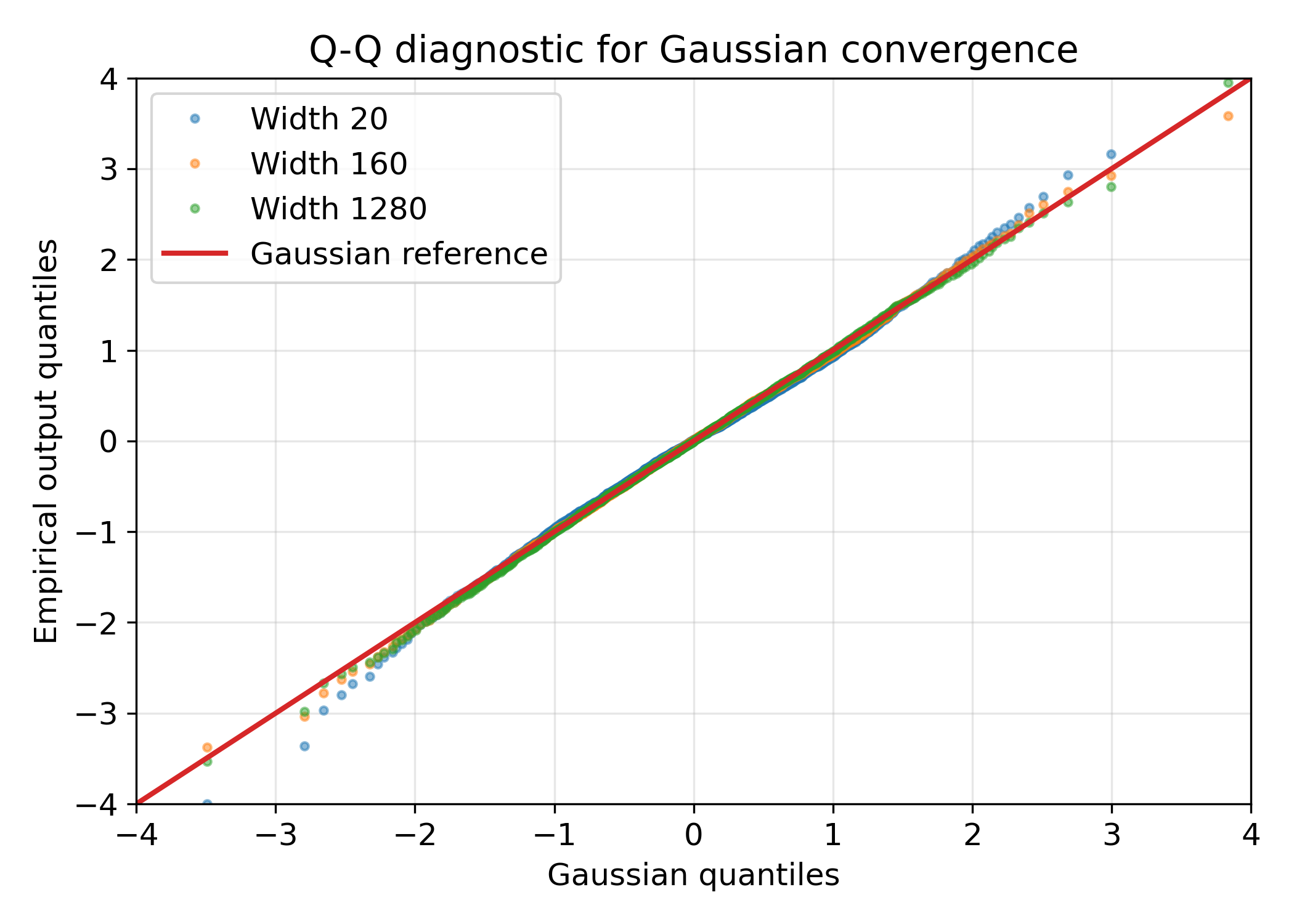}
{Q--Q diagnostic for Gaussian convergence.}
{fig:qq-panel}
\caption{Distribution-level convergence toward Gaussianity. As width increases, the empirical output distribution becomes increasingly consistent with the Gaussian fixed-point description.}
\label{fig:gaussian-diagnostics}
\end{figure}

The approach to Gaussianity can be examined directly at the level of the output distribution. Figure~\ref{fig:gaussian-diagnostics} shows output histograms with increasing widths and gives a Q--Q diagnostic against Gaussian quantiles. The histograms show that the output distribution becomes increasingly consistent with a fitted Gaussian law as width increases. We can clearly see that the larger the width of the output distribution is, the closer to the Gaussian reference line it is. The plots of these histograms are consistent with the central limit interpretation of the infinite-width limit and they support the field-theoretic intuition that the Gaussian process limit is a free fixed point of the induced function distribution and not only an approximation to a particular finite network.

The kernel, cumulant, kurtosis, and distributional diagnostics converge on the same interpretation. The two-point function approaches the infinite-width propagator. Higher-order cumulants are suppressed with width. The one-point output law becomes increasingly Gaussian. These features collectively justify the treatment of the infinite-width limit as an organizing Gaussian fixed point. At the same time, the finite-width corrections remain theoretically significant because they represent the interaction structure that distinguishes a finite neural network from a free Gaussian process. The key part of the theory is the correlation propagation. In mean-field analysis of deep networks, the sensitivity parameter determines whether correlations decay, remain balanced, or amplify with depth. For ReLU activation under symmetric Gaussian preactivations, 
\begin{equation}
\chi
=
\sigma_w^2\mathbb{E}[\phi'(h)^2]
=
\frac{\sigma_w^2}{2}.
\end{equation}
The critical boundary is therefore 
\begin{equation}
\chi=1,
\qquad
\sigma_w=\sqrt{2}.
\end{equation}
We can see that on the ordered side, i.e. $0<\chi<1$, the linearized correlation depth scale is 
\begin{equation}
\xi
=
-\frac{1}{\log \chi}.
\end{equation}
The depth scale diverges as $\chi \rightarrow 1^{-}$. The divergence is similar to the growth of the correlation length near a critical point. In the case of neural networks, it means that correlation structure can persist over more layers as the system approaches the critical boundary \cite{schoenholz2017,poole2016,grosvenor2022}.

\begin{figure}[H]
\centering
\alignedsubfig{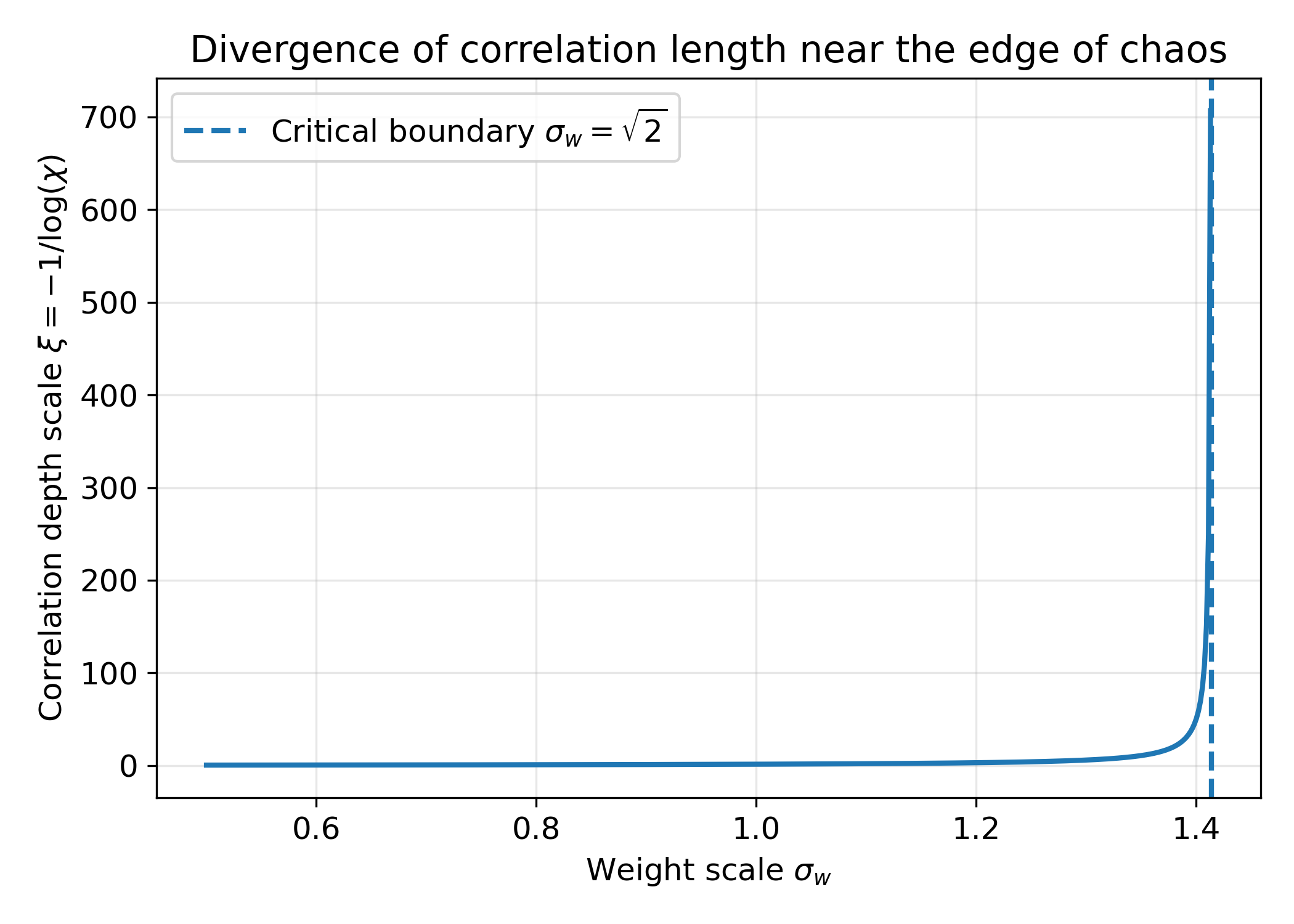}
{Divergence of the correlation depth scale near $\chi=1$.}
{fig:correlation-length-panel}
\hfill
\alignedsubfig{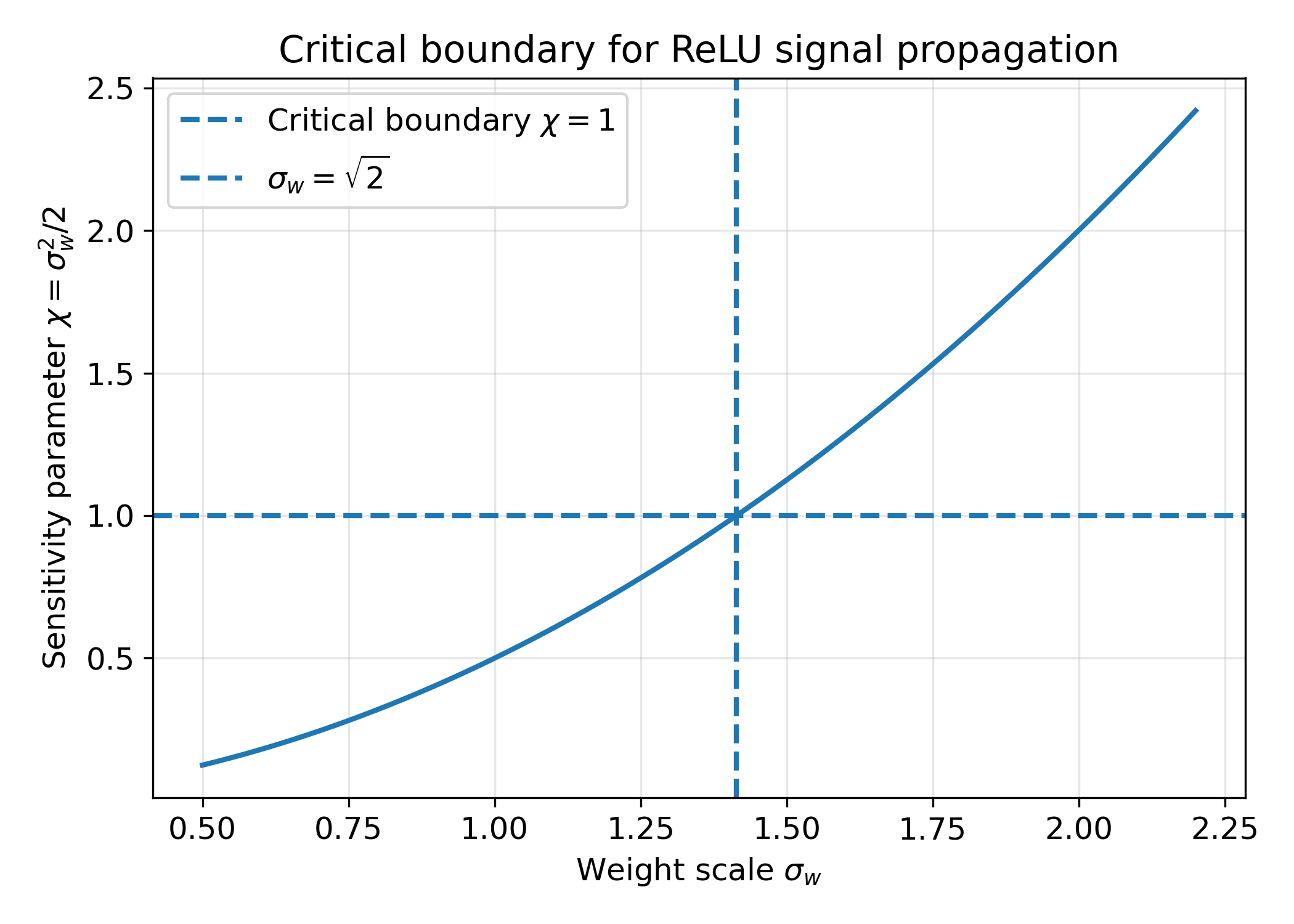}
{ReLU sensitivity boundary $\chi=\sigma_w^2/2=1$.}
{fig:sensitivity-boundary-panel}
\caption{Critical signal propagation near the edge of chaos. The divergence of the depth scale and the crossing of the ReLU sensitivity boundary support the interpretation of criticality as scale-sensitive correlation propagation.}
\label{fig:edge-of-chaos}
\end{figure}

Figure~\ref{fig:edge-of-chaos} shows this critical behaviour. The left panel shows the divergence of the correlation depth scale near the ReLU critical boundary. In the right panel, the sensitivity parameter $\chi=\sigma_w^2/2$ is shown and its crossing is $\chi=1$. These plots relate the edge of chaos condition with the Wilsonian interpretation of criticality. The boundary is not an initialization limit; it is a scale-sensitive area where the description of the effect changes qualitatively. At this surface, the correlation propagation is long-ranged in depth and finite-width effects may be related to the stability and expressivity of the network ensemble \cite{schoenholz2017,poole2016,grosvenor2022}.

\begin{figure}[H]
\centering
\alignedsubfig{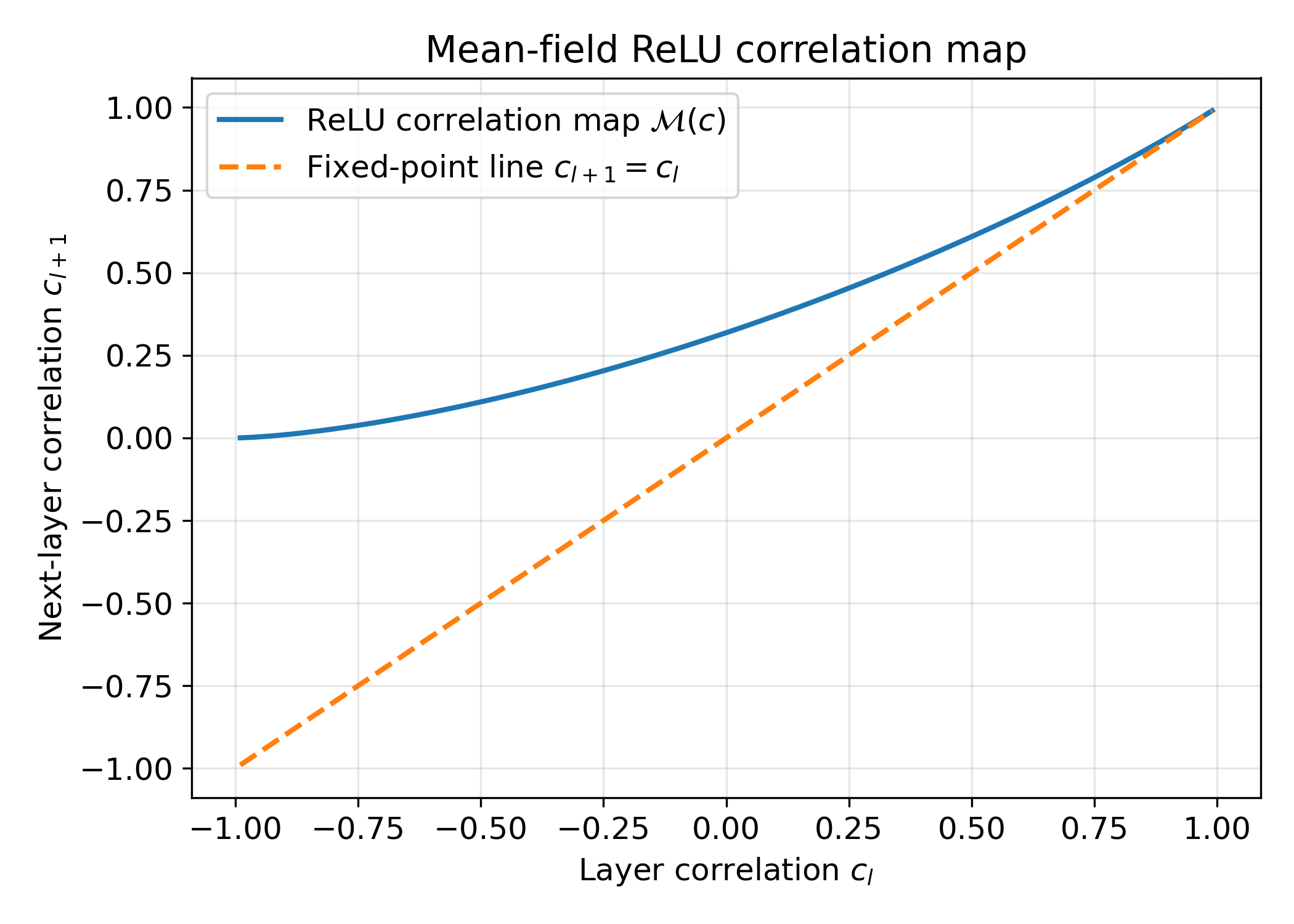}
{Mean-field ReLU correlation map.}
{fig:relu-correlation-map-panel}
\hfill
\alignedsubfig{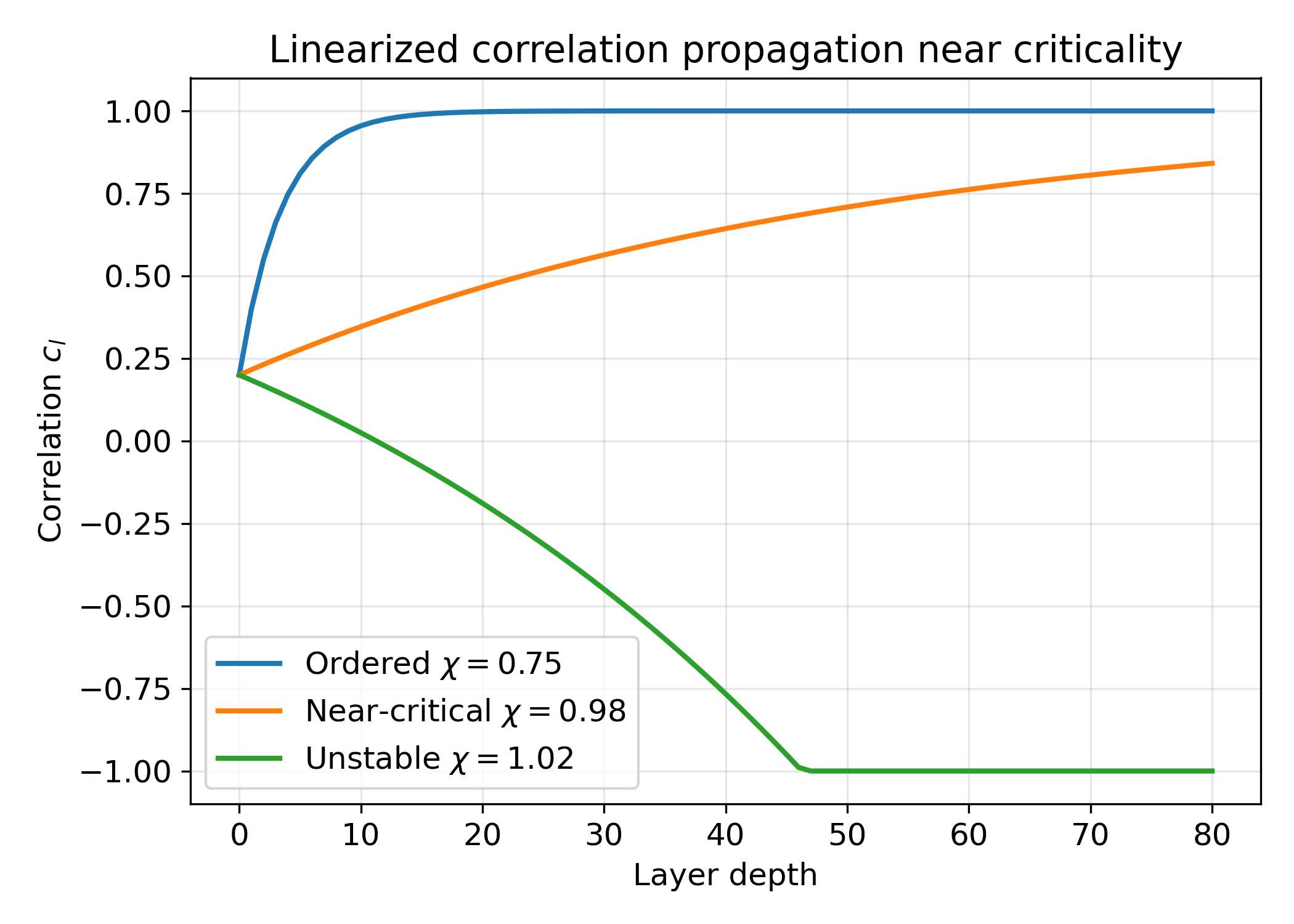}
{Linearized propagation in ordered, near-critical, and unstable regimes.}
{fig:linearized-correlation-panel}
\caption{Correlation dynamics across depth. The mean-field map and linearized propagation show how near-criticality corresponds to slow, scale-sensitive propagation of correlations through layers.}
\label{fig:correlation-maps}
\end{figure}

The layerwise behaviour of correlation propagation is shown in Fig.~\ref{fig:correlation-maps}. The left panel shows the ReLU mean-field correlation map that sends a layer correlation $c_l$ to a next layer correlation $c_{l+1}$. The diagonal line is the fixed point condition $c_{l+1}=c_l$. Near criticality, the map becomes close to tangency with the diagonal and the correlation can evolve slowly over depth. The right panel shows the linearized propagation model for ordered, near-critical and unstable regimes. The ordered regime rapidly collapses correlations, the unstable regime amplifies their perturbations, and the near-critical regime preserves correlation structure over many layers \cite{schoenholz2017,poole2016}.

The correlation-propagation diagnostics also illustrate why criticality can’t be reduced to instability. In the ordered regime, the network loses sensitivity to input differences rapidly because correlations collapse across layers. In the unstable regime, small perturbations are amplified too strongly and make the system difficult to control. The near-critical regime is not like that for both types of systems. It has a slow correlation evolution, long depth scales and a balance between preserving and changing the input structure. That balance is exactly what makes the concept of criticality so useful in this context. Criticality is not associated with maximum chaos but rather with the region where correlations are scale sensitive at depth.

From the Wilsonian perspective, this behavior is naturally associated with marginal or near-marginal perturbations. A strongly irrelevant perturbation disappears very quickly, and it cannot affect the behaviour on the large scale. A strongly relevant perturbation may grow too fast, and it may take the system away from a stable effective description. A marginal or weakly relevant perturbation can still be active on many scales. This is a field-theoretic explanation of why near-critical neural networks can be trained and expressed, in order to preserve structure and not collapse, but are sensitive enough to support nontrivial changes in the input distribution.

The edge-of-chaos plots therefore support the finite-width cumulant results. The cumulant plots describe how the non-Gaussian interaction strength is suppressed with width and correlation propagations describe how information is propagated across depth. Together, the two plots suggest that useful neural behaviour may depend on two types of control: control over finite-width non-Gaussianity and control over depth-wise correlation propagation. A network that is too close to a purely Gaussian and ordered regime may be too rigid and a network with uncontrolled finite-width interactions or unstable propagation may be too sensitive.

These correlation diagnostics provide a functional interpretation of criticality. If correlations collapse too rapidly, distinct inputs cannot be separated out after many layers and representational diversity is lost. If perturbations are large enough, the network is unstable. The near-critical region is located in between these extremes and can be sensitive to scales. In the language of the present paper, criticality is not equal to maximum instability, nor is it equivalent to pure Gaussian simplicity. It is a controlled regime in which correlations remain sensitive to depth and finite-width interactions can remain meaningful without becoming uncontrolled.

The same idea can be formulated in Wilsonian terms. If finite-width effects are perturbations around the Gaussian fixed point, then their importance is determined by their scaling behaviour. Writing the effective action as 
\begin{equation}
S_N[f]
=
S_*[f]
+
\sum_i g_i(N)\mathcal{O}_i[f],
\end{equation}
a Wilsonian transformation causes the coupling flow 
\begin{equation}
\frac{dg_i}{d\ell}
=
\beta_i(\{g\}).
\end{equation}
We may linearize this flow near the fixed point as 
\begin{equation}
\frac{dg_i}{d\ell}
=
y_i g_i
+
\mathcal{O}(g^2).
\end{equation}
The sign of $y_i$ determines the shape of the perturbation: irrelevant perturbations decay, relevant perturbations grow, and marginal perturbations persist at different scales. This classification is important because it distinguishes finite-width corrections that simply disappear from those that determine the effective theory \cite{wilson1974,cardy1996,zinnjustin2002}.

\begin{figure}[H]
\centering
\alignedsubfig{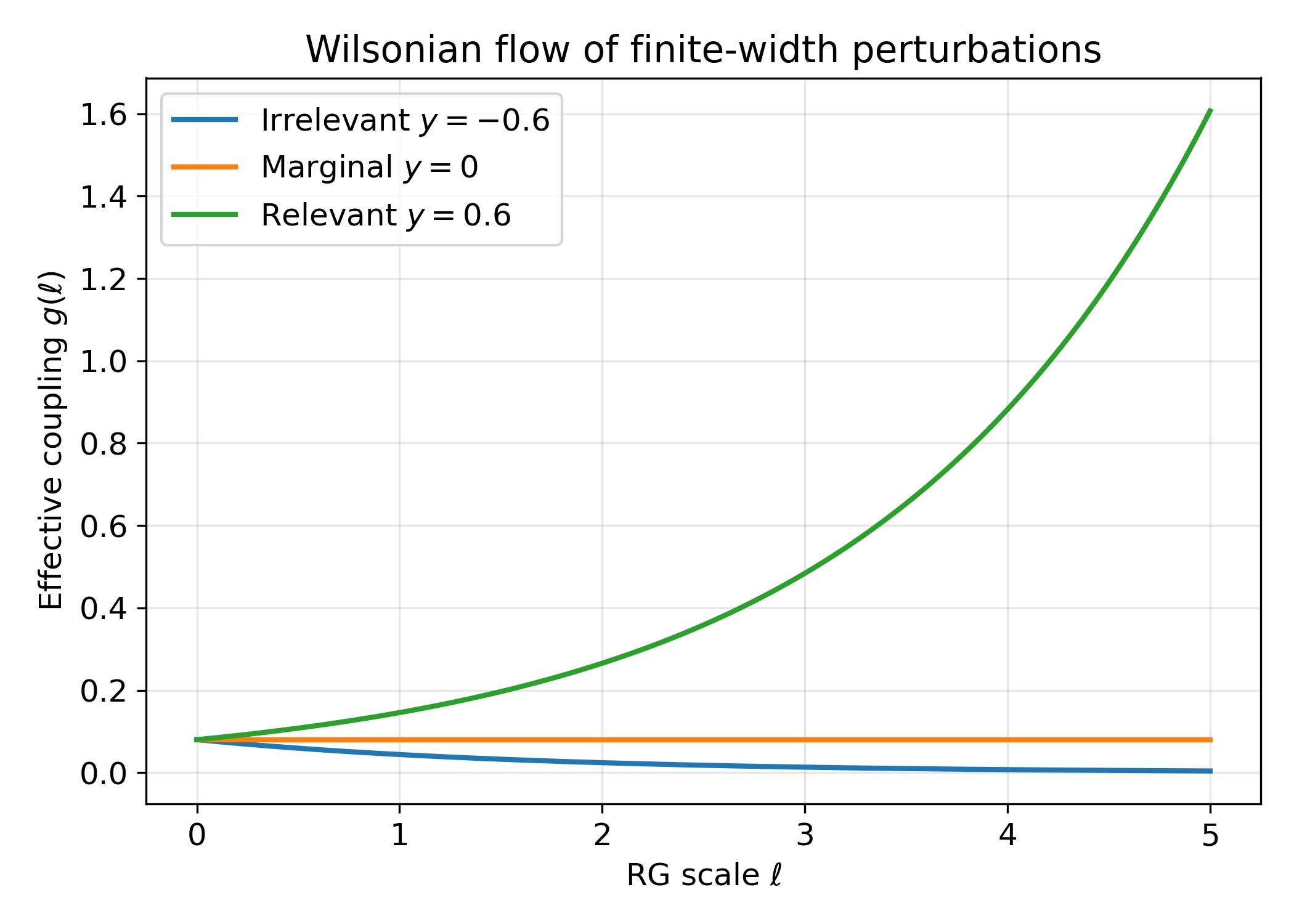}
{Relevant, irrelevant, and marginal coupling flows.}
{fig:wilsonian-flow-panel}
\hfill
\alignedsubfig{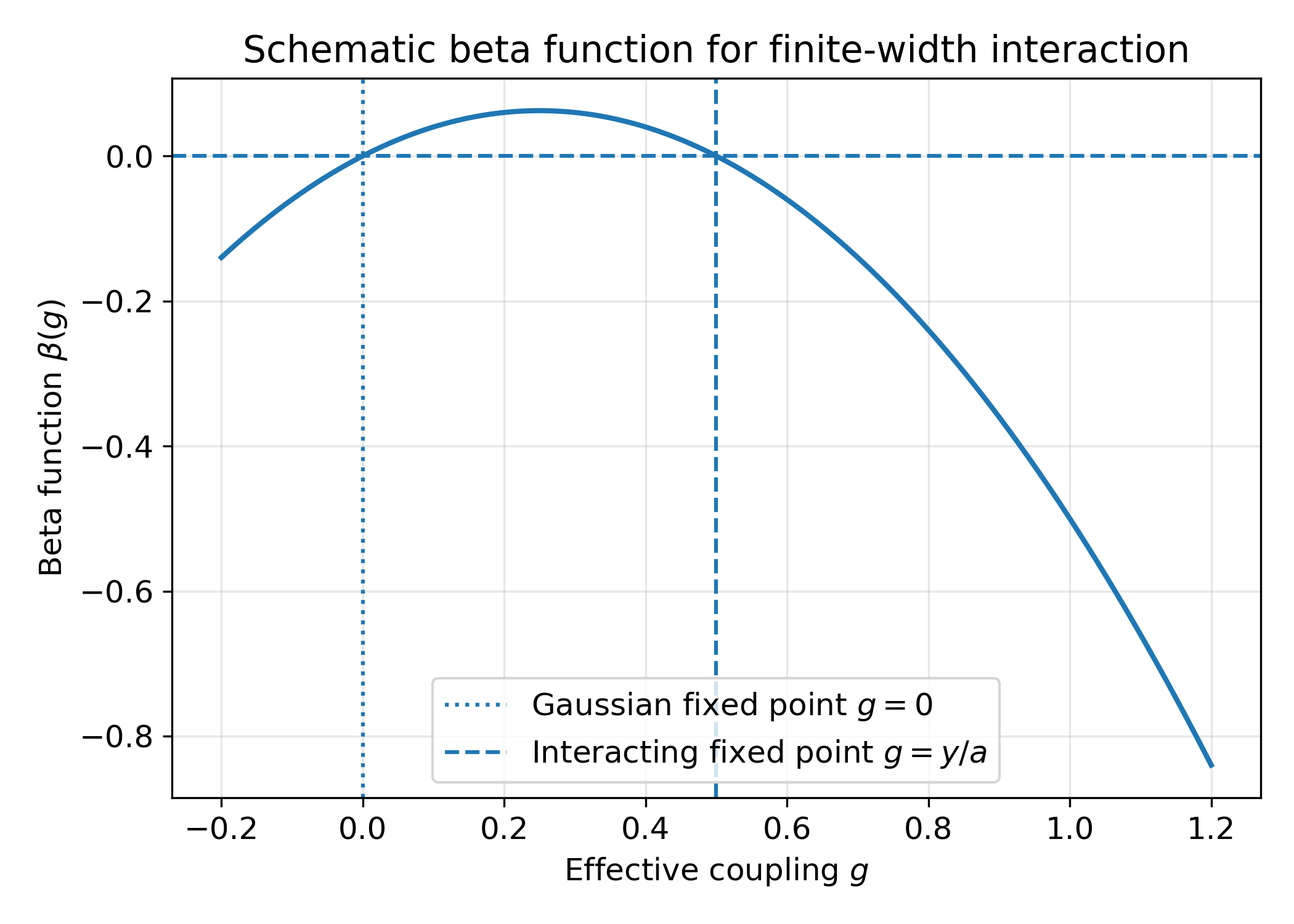}
{Schematic beta function with Gaussian and interacting fixed points.}
{fig:beta-function-panel}
\caption{Wilsonian classification of finite-width perturbations. Finite-width operators may be irrelevant, relevant, or marginal depending on their scaling behaviour near the Gaussian fixed point.}
\label{fig:wilsonian-flows}
\end{figure}

Figure~\ref{fig:wilsonian-flows} shows this classification. The left panel shows the flow of relevant, irrelevant, and marginal couplings. The right panel shows the schematic beta function 
\begin{equation}
\beta(g)
=
yg-ag^2,
\end{equation}
with a Gaussian fixed point at $g=0$ and an interacting fixed point at $g=y/a$. We do not try to establish a specific beta function for each neural architecture in the schematic, but show how the idea of the organizing principle can be illustrated. The finite-width perturbations of small-width perturbations can be different when the effective scale changes. Critical behaviour is associated with perturbations that are active and scale sensitive rather than disappearing immediately or diverging uncontrollably \cite{wilson1974,cardy1996,zinnjustin2002}.

This Wilsonian view gives finite-width corrections a more precise theoretical status. The question is not just whether a neural network has finite width, but which finite-width operators dominate the effective function-space description. A network may have many non-Gaussian corrections, but only some of them may be relevant or marginal to the behaviour of the ensemble. That is the main advantage of the effective-theoretic language: it replaces an undifferentiated notion of finite-width error with a structured classification of perturbations.

Overparameterization can then be reinterpreted in terms of coupling suppression. If finite-width couplings scale as inverse powers of width,
\begin{equation}
g_i(N)
\sim
N^{-\alpha_i},
\qquad
\alpha_i>0,
\end{equation}
then
\begin{equation}
\lim_{N\rightarrow\infty}g_i(N)=0,
\end{equation}
and the finite-width effective action approaches the Gaussian action:
\begin{equation}
\lim_{N\rightarrow\infty}S_N[f]=S_{\infty}[f].
\end{equation}
Thus, increasing width suppresses finite-width interaction terms. In parameter space, this is just a product of increasing the size of the model. In function space, this corresponds to moving to a simpler Gaussian fixed point.

\begin{figure}[H]
\centering
\alignedsubfig{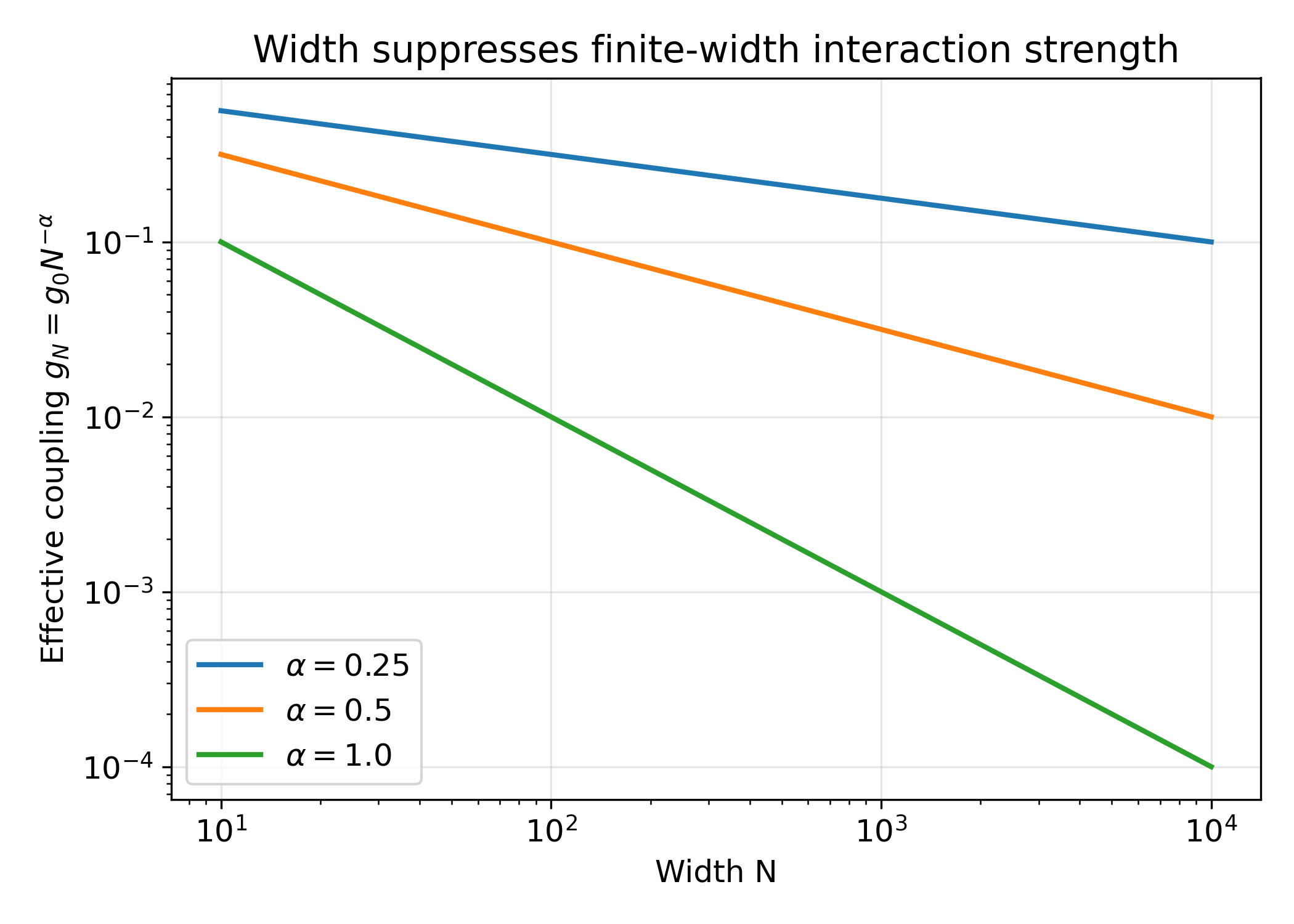}
{Width suppression of effective finite-width couplings.}
{fig:width-suppression-panel}
\hfill
\alignedsubfig{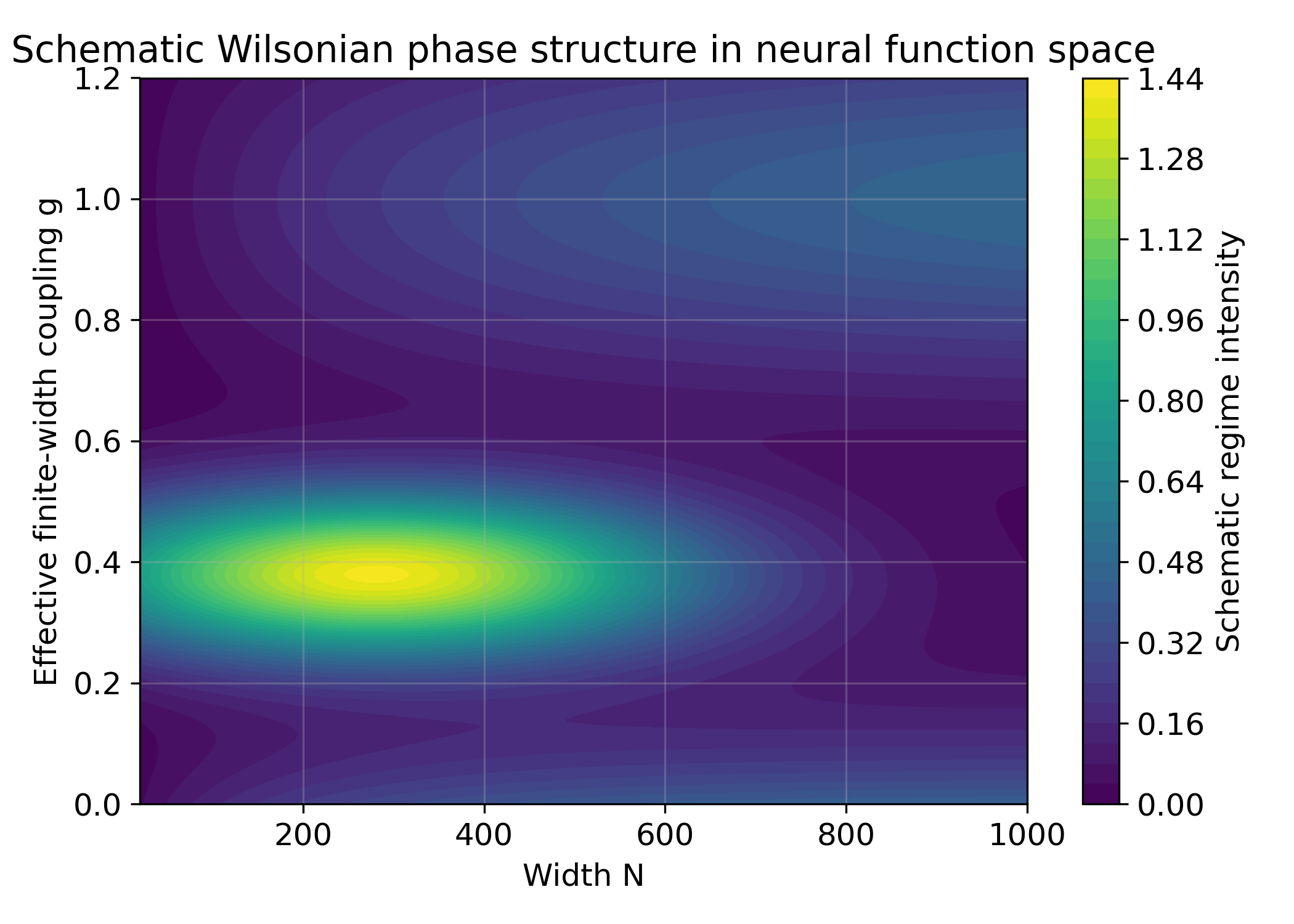}
{Schematic Wilsonian phase structure.}
{fig:phase-diagram-panel}
\caption{Width, interaction strength, and critical structure. Increasing width suppresses effective finite-width couplings, while critical behaviour is represented as a controlled intermediate regime between Gaussian simplicity and strong-interaction instability.}
\label{fig:width-phase}
\end{figure}

As shown in Fig.~\ref{fig:width-phase}, the left panel shows the width-suppressed effective couplings for some scaling exponents. The right panel presents a schematic Wilsonian phase structure in the space of width and effective finite-width coupling. The phase diagram is only a conceptual tool. Its purpose is to summarize the picture of the theory: large width suppresses interaction, too much coupling may lead to instability and non-universality, and critical behaviour is in an intermediate regime of controlled finite-width interaction.

The implication is that overparameterization does not necessarily imply increased complexity of the function space. A highly overparameterized network may have many parameters but a function distribution very close to one based on a Gaussian free theory. On the other hand, a smaller and strongly deformed network may have fewer parameters but stronger non-Gaussian structure. This resolves part of the uncertainty between large number of parameters and stable generalization: the complexity is not simply the number of parameters, but the interaction structure of the induced function space measure.

A clear picture of the theory emerges from all of these diagnostics. Kernel convergence shows how to approach the Gaussian process propagator. Cumulant and kurtosis scaling show that finite-width interaction corrections are suppressed. Distributional diagnostics are convergent to Gaussian output statistics. Correlation lengths and sensitivity plots indicate important behaviour of scale-sensitive propagation. Wilsonian flow diagrams distinguish finite-width perturbations according to their scaling roles. Width-suppression and phase-structure plots show how overparameterization can transfer the efficient theory to Gaussian simplicity and leave room for finite-width criticality.

The collection of diagnostics also suggests a scope of future research. Kernel convergence can be used to assess how quickly architecture approaches its Gaussian process fixed point. Higher-order cumulants can be used to measure the strength of finite-width interaction terms. Output distribution diagnostics can test whether the induced function law is close to Gaussian or still has large non-Gaussian structure. Correlation length estimates can identify the initialization regimes where depth-wise signal propagation becomes important. Coupling flow and phase diagram schematics can then be replaced in future studies by empirically estimated or analytically derived effective beta functions for different architectures. In this way our results not only provide evidence for the theory but also a set of measurable quantities that can be used for future studies of neural criticality.

This is also where the function-space perspective differs from a purely parameter-space perspective. Parameter-space quantities describe the coordinates of the model, but they do not reveal the statistical structure of the functions induced by the model. Function-space diagnostics instead describe the observable distribution of network outputs and their correlations. The kernel, cumulants, output law, correlation length, and effective couplings are thus closer to the objects that determine the behaviour of the neural network as a function approximator. This makes them more suitable for analysing the proposed neural network-quantum field theory correspondence where the central object is not the parameter vector itself, but the ensemble of functions that the network induces.

\section{Discussion}
The Wilsonian approach provides an effective way to unify a number of concepts which are often treated in a separate way in deep learning theory. Infinite width limits, Gaussian processes, finite width corrections, overparameterization, trainability, generalization, edge-of-chaos behaviour, and architectural universality can all be treated in a function space framework. The central idea is that neural networks represent ensembles of functions, and it is possible to analyze these ensembles in terms of effective actions, correlation functions, fixed points, perturbations and scaling behaviour. This makes the language of quantum field theory and statistical mechanics not a sort of loose analogy but rather a mathematical model for how to organise the behaviour of neural network function distributions.

One of the major advantages of this approach is that it clarifies the distinction between parameter-space complexity and function-space complexity. Complexity in deep learning is usually related to the number of parameters or depth of the model. However, the neural network-quantum field theory correspondence shows that a network with a large number of parameters can lead to a relatively simple Gaussian distribution over functions in the infinite width limit, while a finite or trained network can lead to a more complex interacting distribution if its higher-order connected correlations play an important role. This is why classical intuition about the complexity of the model can fail in the overparameterized regime. The number of parameters alone does not determine the structure of the function-space distribution.

Moreover, in many infinite-width approaches finite-width effects are considered to be deviations from the idealized limit. This may make finite-width behaviour seem unnecessary or even impossible. The Wilsonian interpretation we have developed here is an alternative. Finite-width corrections are the terms that lead to interaction, non-Gaussianity, expressivity and potentially critical behaviour. This means that finite-width corrections are not only errors to be eliminated but structures to be classified and understood. This change is important because neural networks are very finite, trained, and architecturally constrained. Their behaviour cannot always be reduced to the Gaussian process limit. 

The new perspective offers a precise way to understand criticality in neural networks. In the literature, criticality in neural networks is generally considered to be a general concept. In this context, we define criticality in terms of the behaviour of connected correlation functions, effective couplings, beta functions and critical surfaces. The notion can be directly related to measurable or computable mathematical objects. A neural network is not close to critical for the reason it is performing well or is large, but for the reason it is scale sensitive in its function space distribution and because finite-width perturbations are feasible or even fine-tunable. This gives neural criticality a better theoretical foundation \cite{halverson2021,erbin2021,grosvenor2022}.

This approach tries to clarify the relation between the edge of chaos and function-space criticality. The edge-of-chaos literature focuses on the propagation of correlations through depth and the balance between ordered and chaotic regimes. The Wilsonian function space view of the edge of chaos is to see the edge of chaos as a critical surface in an effective field theory of network functions. This understanding is not to replace the layerwise dynamical picture, but to embed this in a more general theory of function space distributions. The propagation of correlations through depth is just one part of a larger scaling structure with kernels, connected correlations, finite width operators and effective couplings. The interpretation of training as a deformation of the effective action is also important. Training changes the distribution of function space from a neural network. Gradient descent is not the same as renormalization-group flow, however, and it creates a trajectory in parameter space that changes the effective theory of functions. 

The network could thus be analysed by investigating how its correlation functions and effective couplings differ from those of the initialized network. It is this way to understand generalization and overfitting by studying the structure of trained effective action. A network which generalizes well will preserve the stable relevant structures while suppressing the nonuniversal interactions. A network that overfits might enlarge the data-specific couplings that are not consistent with robust behaviour beyond the training data set. The proposed theory might have implications for architectural comparison. Current deep learning has many architectures, including fully connected networks, convolutional networks, recurrent networks, graph neural networks and transformer-based models. The architectures differ in parameterization, inductive biases, symmetries and computational structure. A field-theoretic approach suggests that they can also be compared by their function space fixed points and finite-width perturbations. Architectures that appear different at the parameter level may belong to the same effective universality classes if they have the same Gaussian limit or the same relevant operators. This opens a way towards a more systematic theory of architectural universality \cite{schoenholz2017,poole2016,grosvenor2022}.

We should note that there are several limitations. The main point is that the meaning of renormalization in neural network function space is not necessarily as simple as in quantum field theory. In physical field theory, renormalization is usually related to changes in length scale in spacetime or momentum space. Neural network input spaces do not always have a natural physical structure or locality structure. Images, text embeddings, graphs, tabular data, and abstract feature spaces all have different notions of distance and scale. Any Wilsonian transformation in function space needs to be clearly defined. The possibilities include coarse-graining through kernel eigenfunctions, data distribution geometry, layerwise correlation structure, spectral decompositions, or functional modes. The present paper does not solve this problem completely, but it identifies it as a key issue for future development \cite{erbin2021,erbin2022}.

The theory is most straightforward for randomly initialized networks, where the ensemble distribution is induced directly from the initialization distribution. Trained networks are more challenging as the distribution of function spaces depends in part on the data, optimization dynamics, the loss functions, the regularization and the training time. A complete theory of trained networks would need a better description of how optimization is transforming the effective action. It might also require nonperturbative approaches, especially if training pulls the network well out of the Gaussian fixed point. 

The paper treats training as a deformation of the effective action, but the exact mathematical form of this deformation is still an open problem. Today's architectures may introduce structures that are not easily captured by simple finite-width expansion. Residual connections, normalization layers, attention mechanisms, gating structures and large-scale pretraining may generate effective interactions whose scaling behaviour is different from those of simpler fully connected networks. This is not to invalidate our Wilsonian view, but it might mean that the effective theory of modern neural networks may require a rich set of operators and symmetries. The classification of such operators will be a question for future work \cite{neal1996,lee2018,jacot2018,erbin2021}. Despite these shortcomings, the Wilsonian function-space interpretation is a promising direction for research. 

Finite-width theory describes perturbations away from that point. Edge-of-chaos theory describes a kind of critical signal propagation. Generalization theory studies stability and transferability of learned functions. The Wilsonian framework, and also that of the network, helps to place these ideas in a common language of effective actions, correlation functions, fixed points, perturbations, and critical surfaces, which is the key to understanding it. The Wilsonian interpretation should be understood as an improvement of the level of analysis. The central object is not the parameter vector but the probability measure over functions generated by the network. In fact, the number of parameters alone does not determine the complexity of a network. A network with many parameters may have a function distribution close to a Gaussian process while a network with a narrow or more trained network may have a stronger non-Gaussian structure. Finite-width effects have a very important theoretical role. They are not just residual errors around the infinite-width limit. They represent the connected higher-order correlations that establish a network that is not just a free Gaussian. While these corrections disappear when the width increases, they may be central to the behaviour of real networks in which width is always finite and training can magnify or modify the terms of interaction.

This interpretation brings criticality into more subtle light. Criticality is not just a loose term for complexity or high performance. It refers to a regime in which correlations, perturbations and effective couplings become sensitive to scale while still being controlled. A purely Gaussian regime is too restrictive and an unstable regime amplifies perturbations too strongly. The critical region lies between these limits, where finite-width structure and correlation propagation are organized and not disappearing or diverging.

We may consider training as a deformation of the induced function-space distribution. Gradient descent is not Wilsonian renormalization but is different in the form of correlations and couplings and thus changes the effective action. Generalization is likely to follow one where the trajectories are stable in the sense that the highly sample-specific interactions are suppressed. Overfitting might correspond to the growth of nonuniversal couplings that describe the training data too closely and does not produce stable behaviour beyond it. This means that future research should look at not only the loss curves and parameter trajectories but also the kernel, connected cumulants, output distributions, correlation length and effective coupling structure before and after training. These quantities would allow us to distinguish useful finite-width interaction from uncontrolled non-Gaussianity and to determine whether a trained network moves toward or away from a controlled critical regime.

\section{Conclusion}

In this paper we have developed a Wilsonian interpretation of criticality in the neural network function space by the neural network -- quantum field theory correspondence. The infinite-width Gaussian process limit has been viewed as a free-field fixed point and finite-width corrections were regarded as non-Gaussian. Connected higher-order correlation functions have been shown to give the characteristic signatures of the departure from the Gaussian limit. For finite-width operators we can treat them as relevant, irrelevant, or marginal depending on their scaling behaviour.

Criticality was defined as a controlled, scale-sensitive regime in which the finite-width interactions and the correlation propagation are significant and not unstable. This is also the explanation for the parameter-space complexity and function-space complexity. Increasing the width of the parameter space may further suppress effective interactions and move the distribution of the functions that are induced into Gaussian simplicity. Training, however, changes the effective action and can move the network to or away from the critical regions for trainability and generalization.

The whole framework is still theoretical and we need to define coarse-graining in neural network function space. Future work should derive effective beta functions, measure connected correlations before and after the training, and investigate nonperturbative regimes where finite-width effects are strong. Learning coarse-graining constructions including holographic machine-learning approaches \cite{howard2021} is one way to define scale-dependent functional degrees of freedom. In conclusion the Wilsonian point of view provides the basis for studying finite neural networks as interacting function-space theories.

\end{document}